\documentclass[a4paper,11pt]{article}
\pdfoutput=1

\usepackage{jheppub}
\usepackage{graphicx,slashed,hyperref,xcolor,multirow}
\usepackage{amssymb}
\usepackage{slashed}
\usepackage{enumitem}
\usepackage{easyReview}
\usepackage{amsmath}
\usepackage{xfrac}
\usepackage{tikz}
\usepackage{comment}
\usepackage[utf8]{inputenc}
\usepackage[compat=1.1.0]{tikz-feynman}
\usepackage{subcaption}
\usepackage[normalem]{ulem}

\def\lsim{\mathrel{\rlap{\lower4pt\hbox{\hskip1pt$\sim$}}
 \raise1pt\hbox{$<$}}}   
\def\gsim{\mathrel{\rlap{\lower4pt\hbox{\hskip1pt$\sim$}}
 \raise1pt\hbox{$>$}}}   

\newcommand{\GeV}{\textrm{ GeV}}

\newcommand{\pppbar}{p p (\bar{p})} 
\newcommand{\zprime}{ {Z^{'}_{B} } } 
\makeatletter

\newcommand{\fmslash}[2][0mu]{%
 \mathchoice
 {\fmsl@sh\displaystyle{#1}{#2}}%
 {\fmsl@sh\textstyle{#1}{#2}}%
 {\fmsl@sh\scriptstyle{#1}{#2}}%
 {\fmsl@sh\scriptscriptstyle{#1}{#2}}}
\newcommand{\fmsl@sh}[3]{%
 \m@th\ooalign{$\hfil#1\mkern#2/\hfil$\crcr$#1#3$}}
\makeatother

\newcommand{\beq}{\begin{equation}}
\newcommand{\eeq}{\end{equation}}
\newcommand{\bea}{\begin{eqnarray}}
\newcommand{\eea}{\end{eqnarray}}
\mathchardef\minus="002D

\newcommand{\met}{{\fmslash E_T}}
\usepackage{xcolor}

\def\beq{\begin{equation}}
\def\eeq{\end{equation}}
\def\bea{\begin{eqnarray}}
\def\eea{\end{eqnarray}}

\title{
``Unification'' of BSM Searches and SM Measurements: the case of lepton$+\met$ and $m_W$
}

\author[a]{Kaustubh Agashe,}
\author[a]{Sagar Airen,}
\author[b]{Roberto Franceschini,}
\author[c]{Doojin Kim,}
\author[d]{Ashutosh V. Kotwal,}
\author[a]{Lorenzo Ricci}
\author[a]{and Deepak Sathyan}

\affiliation[a]{Maryland Center for Fundamental Physics, Department of Physics, University of Maryland,
  College Park, MD 20742, USA}
\affiliation[b]{Universit\`{a} degli Studi and INFN Roma Tre, Via della Vasca Navale 84, I-00146, Rome}
\affiliation[c]{Mitchell Institute for Fundamental Physics and Astronomy, Department of Physics and Astronomy, Texas A\&M University, College Station, TX 77843, USA}

\affiliation[d]{Department of Physics, Duke University, 
Durham, NC 27708}

\emailAdd{kagashe@umd.edu}
\emailAdd{sairen@umd.edu}
\emailAdd{roberto.franceschini@uniroma3.it}
\emailAdd{doojin.kim@tamu.edu}
\emailAdd{ashutosh.kotwal@duke.edu}
\emailAdd{lricci@umd.edu}
\emailAdd{dsathyan@umd.edu}

\preprint{
\begin{minipage}{5cm}
\begin{flushright}
UMD-PP-024-06 \\
MI-HET-831
 \end{flushright}
\end{minipage}
}

\abstract{We develop the idea 
that the unprecedented precision in Standard Model (SM) measurements, with further improvement at the HL-LHC, enables new searches for physics Beyond the Standard Model (BSM).
As an illustration, we demonstrate that the measured kinematic distributions of the $\ell+\met$ final state not only determine the mass of the $W$ boson, but are also sensitive to light new physics. 
Such a search for new physics thus requires a {\em simultaneous} fit to the BSM and SM parameters, ``unifying" 
searches and measurements at the LHC and Tevatron. 
In this paper, we complete the program initiated in our earlier work \cite{Agashe:2023itp}. 
In particular, we analyze ($i$) novel decay modes of the $W$ boson with a neutrinophilic invisible scalar or with a heavy neutrino; ($ii$) modified production of $W$ bosons, namely, associated with a hadrophilic invisible $Z^\prime$ gauge boson; and ($iii$) scenarios without an on-shell $W$ boson, such as slepton-sneutrino production in the Minimal Supersymmetric Standard Model (MSSM). 
Here, we complement our previous MSSM analysis in \cite{Agashe:2023itp} by considering a different kinematic region.
Our results highlight that new physics can still be directly discovered at the LHC, including light new physics,
via SM precision measurements. Furthermore, we illustrate that such BSM signals are subtle, yet potentially large enough to affect the precision measurements of SM parameters themselves, such as the $W$ boson mass.

}

\begin{document}
\maketitle

\section{Introduction}

As we proceed through Run-3 of the LHC and the High Luminosity (HL-LHC) era is just around the corner, we have access to an unprecedented and increasing amount of high-quality data. All this information presents an invaluable stepping stone to a better understanding of nature. Needless to say, we must take advantage of every opportunity to test our knowledge of the Standard Model (SM) and perhaps discover any footprint of new physics Beyond the Standard Model (BSM). 

The conventional approach to new physics (NP) searches involves categorizing data into two distinct regions: $i$) the SM region, where the SM prevails, primarily reserved for measuring SM parameters, and $ii$) the BSM region, characterized by minimal SM backgrounds and therefore optimal target for BSM searches. This approach, however, does not always work. In some cases, the BSM signals populate the SM region and we have no other option than looking for new physics there, where the SM dominates. Even so, we can test the relevant new physics parameter space, provided we have sufficient control of the theory and well-understood data to deal with a small signal-to-background ratio (S/B). Examples of this kind of searches include new physics hiding behind the top decay \cite{Han:2012fw,Czakon:2014fka, Eifert:2014kea, Cohen:2019ycc,Franceschini:2015kdm,Bagnaschi:2023cxg,Franceschini:2023nlp} and di-sleptons hiding behind $WW$ events \cite{Curtin:2013gta,Curtin:2014zua}. Moreover, this paradigm was the core of a recent work \cite{Agashe:2023itp} of ours, where we have shown that the outstanding precision of the recent $W$-boson mass measurements can be repurposed to directly probe new physics in the $\ell+\met$ final state.

\begin{figure}
 \centering
 \includegraphics[width=1\textwidth]{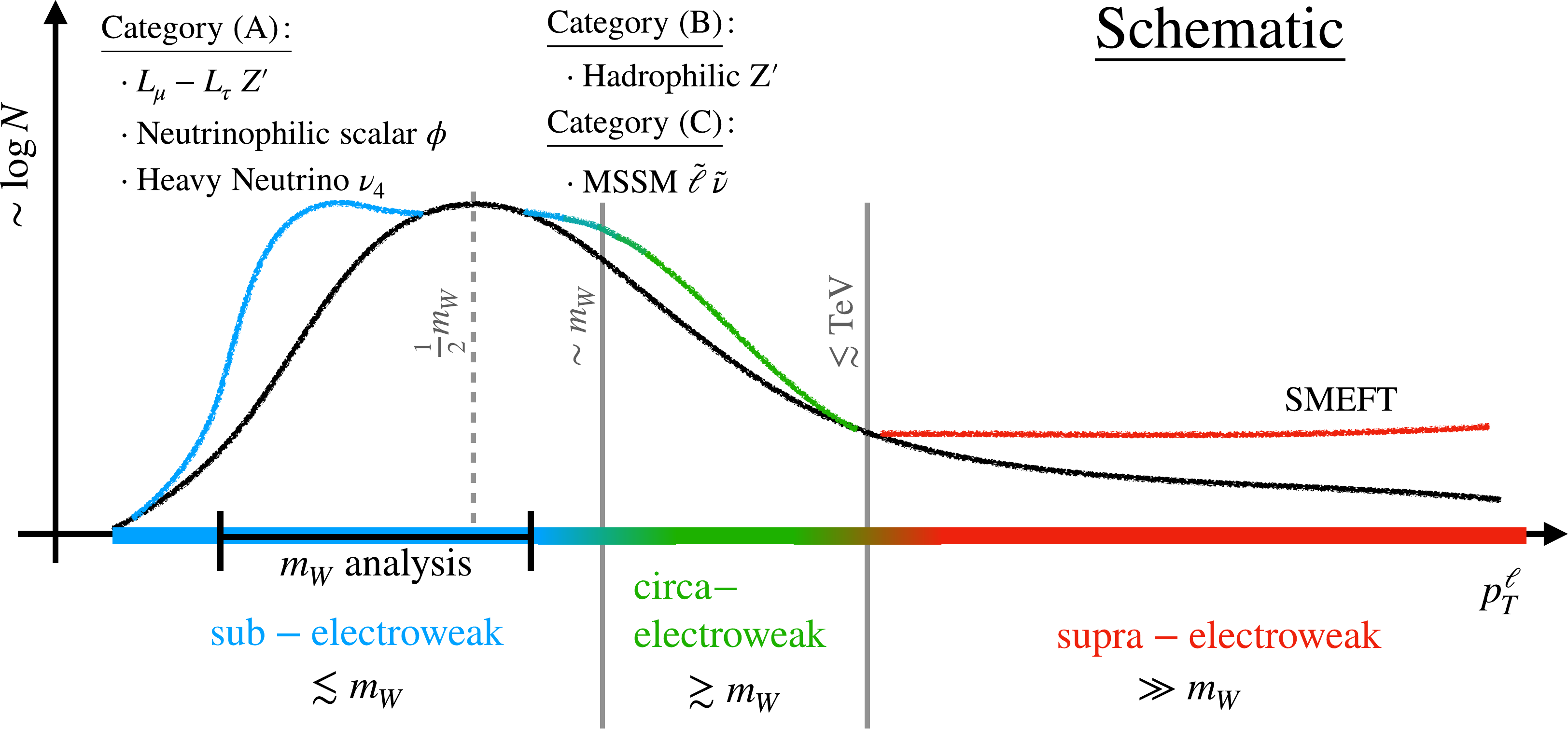}
 \caption{\footnotesize Sketch of the various analysis regions considered in this paper. In black, we show a schematic distribution of the transverse momentum of the lepton from the SM decay of the $W$ boson. The colored lines show roughly where the distribution would change with the inclusion of various BSM physics. The {\bf sub-electroweak} region contains the Jacobian peak~\cite{Smith:1983aa} at $\tfrac{1}{2} m_W$ of the SM $W$ boson distribution, and is therefore used to measure $m_W$: the systematic uncertainties here can reach the per-mil level \cite{ATLAS-CONF-2023-004,ATLAS:2017rzl, CDF:2022hxs}. We proposed in \cite{Agashe:2023itp} that the sub-electroweak region extends the $m_W$ analysis range to further improve sensitivity to BSM. The {\bf supra-electroweak} region contains the high-$p_T$ tails of the distributions and is often used for heavy BSM physics searches, like for $W'$ \cite{CMS:2022krd}, where the systematic uncertainties are usually $\sim 10 \%$. In between these regions, above the Jacobian peak, we identify a {\bf circa-electroweak} region, which holds opportunities for BSM discovery with percent level systematics. The boundaries of the circa-electroweak region are not sharply defined, as they depend on the control of systematics and the sensitivity to new physics. Furthermore, models we considered like hadrophilic $Z'$ and MSSM slepton-sneutrino production can populate both the sub- and circa-electroweak regions.}
 \label{fig:Cartoon}
\end{figure}

In the present paper, we follow this path to complete and extend our analysis in \cite{Agashe:2023itp}. More precisely, taking the case of lepton plus missing transverse energy ($\met$) final state
\begin{equation}
\pppbar \to \ell+\met\,, \label{eq:ellplusmet}
\end{equation}
 we show that the entire (lepton) $p_T$ (and similarly $m_T$) spectrum represents a competitive probe for new physics, not only at the well-exploited TeV region but also down to a few GeV, i.e., closer to the peak of SM events. To represent this situation, we give a sketch of the regions for possible new search strategies in Fig.~\ref{fig:Cartoon}.

In the figure, we define a ``supra-electroweak'' region that covers the high-$p_T$ tail ($\gtrsim$ TeV) of the process Eq.~(\ref{eq:ellplusmet}). This is already the main target of many searches for new particles, e.g., \cite{CMS:2022krd}, and searches for new interactions, see for instance, Refs.~\cite{Farina:2016rws,Torre:2020aiz,Panico:2021vav}. In this regime, systematic uncertainties are generally large,  $\mathcal{O}( 10\%)$, but they benefit from a very favorable S/B ratio. 

On the contrary, in the region of $p_T^\ell$ and $m_T$ where the SM production is copious, we are in the opposite situation: S/B is generally very small, but we can take advantage of remarkable control over not only statistical but also systematic (both theoretical and experimental) uncertainties up to sub-percent level. We call this region of the $p_T^\ell$ and $m_T$ distribution the ``sub-electroweak'' region, i.e., $2 p_T^\ell, \; m_T \lesssim m_W$. The small uncertainties in this region are the result of a large SM production cross-section that allows to have small statistical uncertainties and enables refined calibration methods relying on the abundance of events from SM processes.

The exact extent of the ``sub-electroweak'' region depends on the process and on the observable under examination. In our case, a main driver for the precision of the measured spectra comes from the possibility of using $\pppbar \to Z \to \ell^+ \ell^- $ events to calibrate the predictions of the Monte Carlo codes. Therefore, we can loosely define the sub-electroweak region of the $p_T$ and $m_T$ from zero to the features in these distributions due to the Jacobian peak of Ref.~\cite{Smith:1983aa}.

The sub-electroweak region is suitable for searches for a variety of BSM scenarios, as we already described in \cite{Agashe:2023itp} and as we will show in the rest of the paper. Moreover, as the same data employed for BSM searches are used for the determination of the SM parameters, i.e., $m_W$ in the present study, the search for new physics in the sub-electroweak region requires a {\it simultaneous fit} of the BSM and the SM parameters. Indeed, the case of $m_W$ is a rather rare example of a SM parameter measurement that can be modified directly by the presence of new physics. In some sense, this sort of ``unification'' leads to a possible nontrivial interplay between ``search'' and ``measurement'' activities. 

Finally, we will show that also the region in-between sub-electroweak and supra-electroweak regions, henceforth denoted by the ``circa-electroweak'' region as shown in Fig.~\ref{fig:Cartoon} (i.e., $m_W \lesssim p_T^\ell, \; m_T \lesssim$ TeV), holds a great opportunity for discovery. We will see this through the example of slepton-sneutrino production in the Minimal Supersymmetric Standard Model (MSSM). In particular, we will argue that dedicated measurements in this region, with sufficient control of the systematics at $few$ percent level, can cover unexplored parameter space of the MSSM. Surprisingly, this channel is somehow overlooked by the present literature on slepton searches \cite{ATLAS:2022hbt,CMS:2024gyw}. 

This paper is organized as follows. We start in Sec.~\ref{sec:GenSet} where, partially reviewing our previous work \cite{Agashe:2023itp}, we classify the various BSM scenarios producing the $\ell+\met$ final state into three categories. In this section, we also describe our general procedure and all the technicalities common to the rest of the paper. In Sec.~\ref{Sec:CatA} and Sec.~\ref{Sec:CatB}, we present sensitivity projections for different models whose kinematic distributions populate the sub-electroweak region. 
In particular, in Sec.~\ref{Sec:CatA} 
we discuss two further examples of anomalous $W$ boson decay in category (A), namely, emission of a light invisible scalar from the neutrino and two-body $W$ boson decay involving a heavy neutrino.
Whereas, in Sec.~\ref{Sec:CatB}, we study the new category (B) of anomalous $W$ boson production instead, illustrating it by initial-state radiation of hadrophilic invisible $Z^{ \prime}$ in this process.
In Sec.~\ref{sec:SUSY} we focus on the aforementioned supersymmetry (SUSY) example in the circa-electroweak region, which constitutes the third category (C) of a process with no {\em on}-shell $W$ boson.
This section completes and extends our previous analysis in \cite{Agashe:2023itp} where we considered the same process but focusing only sub-electroweak region. We conclude in Sec.~\ref{Sec:Conclusion}.

\section{Scope and methods of our analysis}\label{sec:GenSet}

\subsection{Classification of New Physics in $\ell+\met$}

\begin{figure}[t!]
  \begin{center}

  \begin{subfigure}[t]{0.45\textwidth}
  \centering
   \begin{tikzpicture}[scale= 0.43]
   \begin{feynman}
   \vertex (a) at (-3,1.5){$q$};
  \vertex (aa) at (3,1.5){$l$};
  \vertex (bb) at (3,-0.8){$\nu_l$};
   \vertex (c) at (-1,0);
   \vertex (cc) at (1,0);
   \vertex (b) at (-3,-1.5){$\Bar{q}$};
   \vertex (d) at (-2,-0.75);
   \vertex (e) at (3,-1.9){$Z'$};
  \diagram*{
  (c)--[boson](cc),
   (a)--[](c),
   (b)--[](c)--[boson,edge label=$W$](cc),
   (aa)--(cc)--(bb),
   (d)--[boson](e),
  };
  \end{feynman};
  \draw [decorate,
   decoration = {brace}] (3.5,-0.7) -- (3.5,-2.1);
   \node at (4.74,-1.4) {\,MET};
  \end{tikzpicture}
  \caption{Hadrophilic $Z'$}
  \label{Fig:Hadrophilic}
 \end{subfigure} 
 \begin{subfigure}[t]{0.45\textwidth}
 \centering
   \begin{tikzpicture}[scale= 0.43]
   \begin{feynman}
   \vertex (a) at (-3,1.8){$q$};
  \vertex (a1) at (3.8,1.8){$l$};
  \vertex (a2) at (3.8,0.8){$\tilde{\chi}_1^0$};
  \vertex (b1) at (3.8,-1.2){$\nu_l$};
  \vertex (b2) at (3.8,-0.2){$\tilde{\chi}_1^0$};
   \vertex (aa) at (2.2,1.3);
  \vertex (bb) at (2.2,-0.7);
   \vertex (c) at (-1,0.3);
   \vertex (cc) at (1,0.3);
   \vertex (b) at (-3,-1.2){$\Bar{q}$};
  \diagram*{
  (c)--[boson](cc),
   (a)--[](c),
   (b)--[](c)--[boson,edge label=$W^*$](cc),
   (aa)--[scalar,edge label' = $\tilde{l}$](cc)--[scalar,edge label' = $\tilde{\nu}$](bb),
   (aa)--(a1),
   (aa)--[boson](a2),
   (aa)--(a2),
   (bb)--(b1),
   (bb)--[boson](b2),
   (bb)--(b2),
  };
  \end{feynman};
  \draw [decorate,
   decoration = {brace}] (4.4,1.1) -- (4.4,-1.5);
   \node at (5.5,-0.1) {\, MET};
  \end{tikzpicture}
  \caption{MSSM slepton-sneutrino}
  \label{Fig:MSSM}
 \end{subfigure}
 
 \vspace{1cm}
 \begin{subfigure}[t]{0.45\textwidth}
 \centering
  \begin{tikzpicture}[scale= 0.43]
   \begin{feynman}
   \vertex (a) at (-3,1.5){$q$};
  \vertex (aa) at (3,1.5){$l$};
  \vertex (bb) at (3,-1.5){$\nu_l$};
   \vertex (c) at (-1,0);
   \vertex (cc) at (1,0);
   \vertex (b) at (-3,-1.5){$\Bar{q}$};
   \vertex (d) at (2,-0.75);
   \vertex (e) at (3,0.25){$\phi$};
  \diagram*{
  (c)--[boson](cc),
   (a)--[](c),
   (b)--[](c)--[boson,edge label=$W$](cc),
   (aa)--(cc)--(bb),
   (d)--[scalar](e),
  };
  \end{feynman};
  \draw [decorate,
   decoration = {brace}] (3.5,0.2) -- (3.5,-1.8);
   \node at (4.7,-0.7) {\, MET};
  \end{tikzpicture}
  \caption{Neutrinophilic scalar}
  \label{Fig:NeutrinophilicScalar}
 \end{subfigure}
 \begin{subfigure}[t]{0.45\textwidth}
 \centering
  \begin{tikzpicture}[scale= 0.43]
   \begin{feynman}
   \vertex (a) at (-3,1.5){$q$};
  \vertex (aa) at (3,1.5){$l$};
  \vertex (bb) at (3,-1.5){$\nu_4$};
   \vertex (c) at (-1,0);
   \vertex (cc) at (1,0);
   \vertex (b) at (-3,-1.5){$\Bar{q}$};
  
  \diagram*{
  (c)--[boson](cc),
   (a)--[](c),
   (b)--[](c)--[boson,edge label=$W$](cc),
   (aa)--(cc)--(bb),,
  };
  \end{feynman};
  \draw [decorate,
   decoration = {brace}] (3.5,-1.05) -- (3.5,-1.8);
   \node at (4.7,-1.3) {\,MET};
  \end{tikzpicture}
  \caption{Heavy neutrino}
  \label{Fig:HeavyNeutrino}
 \end{subfigure}
 \end{center}
 \caption{New physics contributions to $\ell+\met$ for the different models that we consider.\label{Fig:FeynDiag}}

 \end{figure}
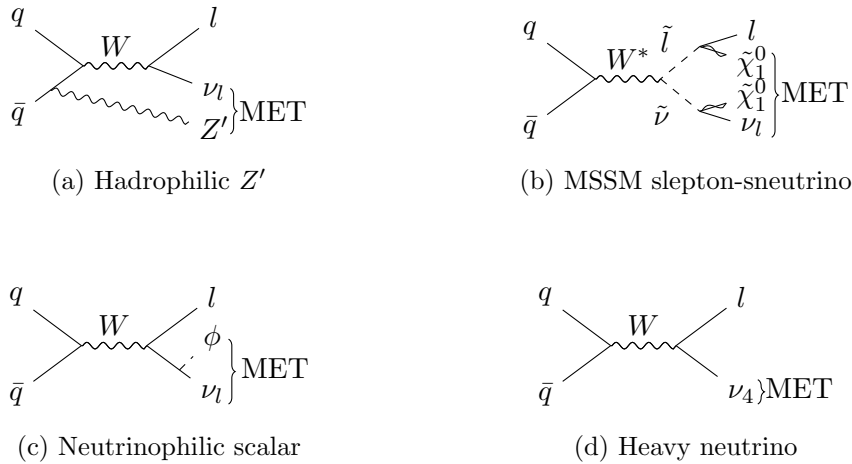

As mentioned in the introduction, this basic signal is already the main target of a plethora of BSM searches: from new ``on-shell'' particles, which can be directly produced, to heavy new physics, which can only be tested indirectly through new interactions within the SM. A model-independent analysis that encompasses all this new physics would definitely be interesting, but it is beyond the scope of this paper, at least for the direct production of new particles.
Given that the case of heavy new physics affecting the supra-electroweak region of the kinematic distributions is already covered extensively (e.g., in Refs.~\cite{Farina:2016rws,CMS:2022krd}), we proceed to classify all the relevant sub-electroweak and circa-electroweak new physics into three categories (\cite{Agashe:2023itp}):

\begin{enumerate}[label=(\Alph*)]
 \item \textbf{Anomalous $W$-boson decay}\newline
 The first category collects all the new physics models where the decay of the $W$ boson is different from the SM
 \begin{align}
  \pppbar \rightarrow W \rightarrow \ell + X \,(\text{invisible})\,,
 \end{align}
 where $X$ stands for the set of (collider-)invisible final-state particles.
 We investigated a well-motivated example of this sort in our previous work \cite{Agashe:2023itp}: the so-called $L_{\mu}-L_{\tau}$ gauge boson $Z'$. Under this scenario, the $W$ boson is allowed to have an exotic three-body decay mode, namely $X\equiv \nu_{\mu} + Z'$ with $Z'$ possibly decaying further into an invisible final state. We will study two more examples belonging to this category in Sec.~\ref{Sec:CatA}: a new boson radiated from SM neutrinos and a fourth neutrino coupled to the $W$ boson, as displayed respectively in Figs.~\ref{Fig:NeutrinophilicScalar} and \ref{Fig:HeavyNeutrino}.
 As we have shown in \cite{Agashe:2023itp} and as we will show in Sec.~\ref{Sec:CatA}, in general, new physics of this sort renders the $p_T$ spectrum of the lepton softer, populating the low $p_T$ region below the $m_W/2$ peak (see for instance, Fig.~\ref{Fig:Dist}). As a result, models in category (A) are studied in the sub-electroweak region as shown in Fig.~\ref{fig:Cartoon}, using the same parameter space as the $m_W$ measurement, with proposed extensions to lower $p_T^\ell$ and $m_T$ to improve sensitivity to new physics.
  
 \item \textbf{Anomalous $W$-boson production}

 The second possibility is to modify the (single) $W$ production to the one in association with new invisible particle(s)
 \begin{align}
  p p (\bar{p}) \rightarrow X(\text{invisible})+ W(\to \ell\, \nu_l)\,.
 \end{align}
 In this case, the final state consists of a real (on-shell) SM $W$ boson and extra invisible particle(s). A reference Feynman diagram for this category is displayed in Fig.~\ref{Fig:Hadrophilic}.
 We will study this category in Sec.~\ref{Sec:CatB} through the example of an invisible $Z'$ coupled to quarks (see for instance, Ref.~\cite{Dobrescu:2021vak}). Here $X\equiv Z'$ with the $Z'$ produced in association with the $W$. Another study in a similar direction can be found in \cite{Bandyopadhyay:2022bgx}.
 In general, in this category, $X$ recoils against the $W$, making the $p_T^\ell$ spectrum harder than in pure SM production, as shown, for instance, in Fig.~\ref{Fig:Dist}. As a result, models in category (B) can populate both the sub- and circa-electroweak regions as shown in Fig.~\ref{fig:Cartoon}.
 
 \item \textbf{$\ell+\met$ not from an on-shell $W$ boson}

 Finally, the third possibility is that new physics directly mediates the production of $\ell+\met$, without the production of any on-shell $W$ boson. 
 A simple but well-motivated example arises in any models featuring an extra charged gauge boson (often denoted by $W'$) decaying into $\ell \nu$ \cite{Barger:1980ix,Pappadopulo:2014qza,ATLAS:2024qvg,CMS:2019kaf}. Another example is $\ell+\met$ from slepton-sneutrino production in the MSSM, as in Fig.~\ref{Fig:MSSM}, which was already discussed in \cite{Agashe:2023itp} and will be further analyzed in Sec.~\ref{sec:SUSY}. In addition, new interactions among the SM particles, e.g., new dimension-6 quark-lepton interactions, also belong to this category. 
 Unlike the previous two categories, there is no generic expectation for the $p_T$ and $m_T$ spectra, which may change from model to model. As the region for supra-electroweak momenta is well covered by new physics searches at the LHC experiments, in the spirit of our paper to pursue BSM signals in SM-rich regions of parameter space, we will focus on sleptons contributing to the sub-\footnote{The sub-electroweak region has been already studied by us in \cite{Agashe:2023itp}.} and circa-electroweak $p_T$ regions (see Figs.~\ref{fig:Cartoon} and \ref{fig:SUSY_dists_WPrime}). The slepton masses of interest are just above the LEP bounds and in some cases are not yet probed by the LHC.
\end{enumerate}

While we believe that this classification is rather comprehensive, it is not intended to describe all of the phenomenology; various new physics models in a given category can give rise to different signals and may require distinct dedicated analyses. We believe, though, that through the various aforementioned examples, we are able to capture the qualitative phenomenological features of all the new physics in $\ell+\met$ not covered by the majority of the literature, that mostly deals with supra-electroweak scale $p_T$.

\subsection{Sensitivity to new physics}\label{subsec:technicalities}
Our sensitivity projections are based on analyses of kinematic distributions. As to profit as much as possible from measurements that have been {\it already} carried out, we choose the distributions of lepton $p_T^\ell$ and the transverse mass $m_T$ of the $\ell+\met$ system for our LHC projections. We take as reference the $W$ boson mass measurements from ATLAS~ \cite{ATLAS:2017rzl,ATLAS-CONF-2023-004}. For CDF projections, we also include the missing transverse momentum $p_T^{\text{miss}}$ distribution, as it was done by the experiment itself in their $W$ boson mass measurement~\cite{CDF:2022hxs}. As pointed out in \cite{Agashe:2023itp}, there is a remarkable interplay between new physics searches and SM measurements, so that the two must be considered part of the same analysis. Concretely, this entails that the interpretation of the data in these models requires the simultaneous fit of $m_W$ and the new physics parameters. Analogously, a fair SM measurement for $m_W$ should take into account possible BSM effects as nuisance parameters in the likelihood. 

We thus construct a $\chi^2$ for each model and for each binned distribution of the observable $\mathcal{O}$ 
\begin{align}
 \chi^2_{\mathcal{O}}(\Delta_{m_W} , \theta_{\text{NP}}) = \sum_{i,\,j =1}^{N_{\text{Bins}}} \left(N^i(\Delta_{m_W} , \theta_{\text{NP}})-\overline{N}^i \right)\Sigma^{-1}_{ij}\left(N^j(\Delta_{m_W} , \theta_{\text{NP}})-\overline{N}^j \right)\,,
 \label{Eq:ChiSqu1}
\end{align}
where $\bar{N}^i$ is the observed number of events and $N^i (\Delta_{m_W}, \theta_{\text{NP}})$ is the expected number of events for observable $\mathcal{O}$ in the bin $i$ as a function of $m_W$. The actual value of $m_{W}$ will not matter, since we track the $m_{W}$ dependence in our calculations through the mass difference $\Delta_{m_W} = m_W - \overline{m}_W$, where $\overline{m}_W$ is the $W$ mass in the MC sample that we use as a substitute for the data from the actual experiment. Notably, our $\chi^{2}$ depends on the new-physics (NP) parameters $\theta_{NP}$, whose nature will differ for each of the models under consideration in the following.
Here, $\Sigma_{ij}^{-1}$ denotes the inverse of the covariance matrix $\Sigma_{ij}$ including various uncertainties
\begin{equation}
 \Sigma_{ij} = \Sigma_{ij}^{\text{stat}}+\Sigma_{ij}^{\text{unc}}+\Sigma_{ij}^{\text{cor}}\,.
\end{equation}
The first term accounts for the statistical uncertainties while the second and the third terms model the systematic uncertainties through completely uncorrelated and completely correlated components, respectively, common for all the bins:
\begin{align}
 \Sigma_{ij}^{\text{stat}} = \overline{N}^i \delta_{ij}\,, && \Sigma_{ij}^{\text{unc}} = (\overline{N}^i \epsilon^{\rm unc} )^2\delta_{ij} \,, && \Sigma_{ij}^{\text{cor}} = \overline{N}^i\overline{N}^j (\epsilon^{\rm cor})^2 \,,
\end{align}
where $\delta_{ij}$ is the usual Kronecker delta while $\epsilon^{\rm unc}$ and $\epsilon^{\rm cor}$ are fractional factors parameterizing the uncorrelated and correlated systematics, respectively.
The pseudo-data sets are generated assuming the SM with $W$-boson mass $\overline{m}_W$, hence the $\chi^2$ is minimized at $\Delta_{m_W}= \theta_{NP} = 0$. 

As anticipated in the introduction, we will perform two different types of analyses targeting sub-electroweak scale new physics or circa-electroweak scale new physics. For these targets we will adopt, respectively, two baseline analysis setups: 
$i$) the sub-electroweak analysis described in Sec.~\ref{sec:subEW}, in which we try as much as possible to study distributions obtained under event selection closely following the ATLAS and CDF $W$ boson mass measurements; 
$ii$) the circa-electroweak analysis, described in Sec.~\ref{sec:circaEW}, in which we allow for larger deviations in the event selection, and in particular, in the range of $p_{T,\ell}, m_T$, and $p_T^{\text{miss}}$, as to maximize the reach in the search for new physics. 

All our LHC projections are based on Monte Carlo (MC) simulations produced via \textsc{MadGraph5\_aMC@NLO}v3.42 \cite{Alwall:2014hca} + \textsc{PYTHIA8.212} \cite{Bierlich:2022pfr} + \textsc{Delphes}v3.4 \cite{deFavereau:2013fsa} (ATLAS card). Regarding the parton distribution functions (PDFs), we employ \textsc{LHAPDF} \cite{Buckley_2015} with PDF ID:244800 \cite{Ball_2013}. Pileup of collisions, with average number of pileup events per bunch crossing $\langle \mu \rangle$ = 50, is simulated through the dedicated \textsc{Delphes} ATLAS card. In our results, pileup is always included unless indicated otherwise. For the CDF projections, we simulate detector effects through a custom MC code package, implementing the results of \cite{CDF:2022hxs} (see Appendix of Ref.~\cite{Agashe:2023itp} for details). The rest of our emulation of the CDF detector matches our emulation of the ATLAS detector. For concreteness, we only simulate the cases resulting in muons in the final state. Nevertheless, we expect that completely analogous results will apply to the cases involving electrons in the final state. In the following, we analyze the muon final state without performing any combination with the electron final state unless noted otherwise.

We do not include any backgrounds to the $W$ boson mass measurement in the $\chi^2$ of Eq.~\eqref{Eq:ChiSqu1} because these backgrounds can be estimated from MC and are only a few percent~\cite{ATLAS:2017rzl,ATLAS-CONF-2023-004,CDF:2022hxs} of the SM $W$ boson events in our analysis range. Indeed, for our analyses in the sub-electroweak region (in Sec.~\ref{Sec:CatA} and Sec.~\ref{Sec:CatB}), we consider analysis ranges similar to those of ATLAS \cite{ATLAS:2017rzl, ATLAS-CONF-2023-004} and CDF \cite{CDF:2022hxs}; we slightly enlarge the ranges to maximize the sensitivity to new physics (see Tab.~\ref{tab:CUTS}). On the other hand, for our analysis in Sec.~\ref{sec:SUSY}, we substantially modify the analysis range from the $W$-boson mass measurement (see Tab.~\ref{tab:susy_cuts}) to the circa-electroweak region. Yet, even in this case, we expect backgrounds to be only marginally relevant, as we comment in detail in Sec.~\ref{sec:SUSY}.

 \subsubsection{Sub-electroweak analysis \label{sec:subEW}}
 As shown in our earlier work \cite{Agashe:2023itp} and expanded in later Sec.~\ref{Sec:CatA} and Sec.~\ref{Sec:CatB}, a great deal of the sensitivity to new physics comes from the $p_T$ and $m_T$ spectra near the SM Jacobian peak~\cite{Smith:1983aa}. Therefore, we design our analysis as much as possible around the currently published $W$-boson mass measurement \cite{ATLAS:2017rzl,ATLAS-CONF-2023-004, CDF:2022hxs}, just slightly extending the range for improved sensitivity to new physics. For reference purposes, we give in Tab.~\ref{tab:CUTS} a summary of the event selection used for the current measurements of the $W$ boson mass and in our sub-electroweak analyses. 
 
 For the sub-electroweak analyses, we only use the shape of the distribution for the search for new physics and the simultaneous determination of $m_W$. The same approach is used by ATLAS in their $W$ mass measurements. Therefore, we normalize the expected number of events $N_{ev}^i$ in Eq.~\eqref{Eq:ChiSqu1} to the pseudo-data in each computation of the $\chi^2$ between pseudo-data and the expected distribution. 
 Because of this normalization, we do not include any correlated components in the systematics (i.e., $\epsilon^{\rm cor} = 0$).
 For the uncorrelated components, we consider a few benchmark scenarios; we consider $\epsilon^{\rm unc}=0\%,\, 0.1\%$, and 0.5\% for the LHC projections and $\epsilon^{\rm unc}=0\%$ and $1\%$ for the CDF ones.

 \subsubsection{Circa-electroweak analysis \label{sec:circaEW}}
 As will be discussed in detail later in Sec.~\ref{sec:SUSY}, some sensitivity to BSM comes from the $p_T$ range beyond the SM Jacobian peak~\cite{Smith:1983aa}. Therefore, we consider the possibility to perform the search for BSM physics in a different range than what was considered in the measurement of $m_W$. We focus, in particular, on new physics appearing at a mass scale around the weak scale, and thus, we denote the associated region and strategy as ``circa-electroweak.'' 
 
We will use the example of sleptons in the MSSM immediately above the LEP limits to display one instance of this type of search. In this context, we will illustrate how the proposed search strategy can fill the gaps in current searches.
 
The analysis has some slight differences compared to the ``sub-electroweak'' analysis. First of all, it differs by the range of kinematic quantities that we use in our likelihood computations from SM templates and pseudo-data from BSM signals. For illustration of the selection we will use, we refer the reader to Tab.~\ref{tab:susy_cuts}.
Furthermore, in this analysis, part of the sensitivity stems from the overall number of events. Thus, unlike for the sub-electroweak analysis, we do not normalize the predicted SM distributions to the pseudo-data. For this reason, we include a correlated error $\epsilon^{\rm cor}= 2\%$, a conservative estimate based on LHC Run-2 analyses \cite{ATLAS:2022hbt,CMS:2023qhl}. We consider a few benchmark points for the uncorrelated per-bin systematics: $\epsilon^{\rm unc}=0\%,\, 2\%,\, 5\%$, and 10\% for the LHC projections. Unlike for the sub-electroweak analysis, the CDF projections are not included, as the final-state particles are heavy enough ($m_{\tilde{\ell}}\gtrsim 100$ GeV) to render the cross-sections at CDF negligible.
 Finally, a very important point is that in the circa-electroweak analysis, the exact value of the $W$-boson mass is not a crucial input. This is because the majority of the sensitivity to new physics comes from regions of the kinematic distributions in which the SM processes are relatively rare, thus a moderately large S/B is typical for this search strategy. Thus, after careful testing of this step, we assume $m_W$ to be a known parameter from other experiments or other analyses, or both, and set $\Delta m_W = 0$ in Eq.~\eqref{Eq:ChiSqu1}.

\begin{table}[]
 \centering
 \resizebox{\columnwidth}{!}{
 \begin{tabular}{c||c|c|c|c|c|c}
  &$p_T^{\ell}$ & $p_T^{\text{miss}}$ & $m_T$ & $|\vec{u}_T|$& $m_T$ range &$p_T^{\ell}$ range \\ \hline\hline
 \parbox{3cm}{\centering ATLAS \cite{ATLAS:2017rzl,ATLAS-CONF-2023-004} \\($W \rightarrow \mu\, \nu_{\mu}$)} & \parbox{2.5cm}{$> 30$ (analysis) \\ $> 18$ (trigger)} & $>30$ & $> 60$& $<30$ &$[60,100]$ & $[30,50]$\\ \hline
 Sec.~\ref{Sec:Neut} ($W\rightarrow \mu \nu_{\mu} \phi$) & $> 20$ &$>20$ &$>40$& $<30$& $[40,100]$&$[20,50]$ \\ \hline
 Sec.~\ref{Sec:Massv} ($W\rightarrow \mu \nu_{4}$)&$> 20$ &$>20$ &$>40$& $<30$& $[40,100]$&$[20,50]$ \\ \hline 
 Sec.~\ref{Sec:CatB} ($pp \rightarrow W Z'$) &$> 30$ &$>30$ &$>60$& $<30$& $[60,140]$&$[30,70]$ \\ \hline \hline
 CDF ($\mu$) \cite{CDF:2022hxs} & \parbox{3cm}{$[30, 55]$ (analysis) \\ $>18$ (trigger)}& $[30,55]$& $[60,100]$ & $<15$& $[65,90]$ & $[32,48]$ \\ \hline 
 Sec.~\ref{Sec:Neut} ($W\rightarrow \mu \nu_{\mu} \phi$) & $[20,55]$& $[20,55]$& $[60,100]$ &$<15$&$[40,90]$& $[20,48]$\\ \hline
 Sec.~\ref{Sec:Massv} ($W\rightarrow \mu \nu_{4}$)& $[20,55]$& $[20,55]$& $[60,100]$ &$<15$&$[40,90]$& $[20,48]$
 \end{tabular}
 }
 \caption{\footnotesize Kinematic cuts and analysis ranges considered in our fit and in the latest $W$-mass measurements \cite{ATLAS:2017rzl,ATLAS-CONF-2023-004,CDF:2022hxs}. All the number are measured in GeV. The hadronic recoil vector is denoted by $\vec{u}_T$. For our LHC projections, we construct $2$ GeV bins for $m_T$ and $1$ GeV bins for $p_T^\ell$ \cite{ATLAS-CONF-2023-004}, unless otherwise specified. For the CDF projections, we construct $0.5$ GeV bins for $m_T$ and $0.25$ GeV bins for $p_T^\ell$, $p_T^{\text{miss}}$ \cite{CDF:2022hxs}. The analysis ranges of $p_T^{\ell}$ apply also to $p_T^{\text{miss}}$ in the CDF analyses.}
 \label{tab:CUTS}
\end{table}

\begin{figure}[h]
\centering
\begin{minipage}{0.48\textwidth}
\centering
\includegraphics[width=1\textwidth]{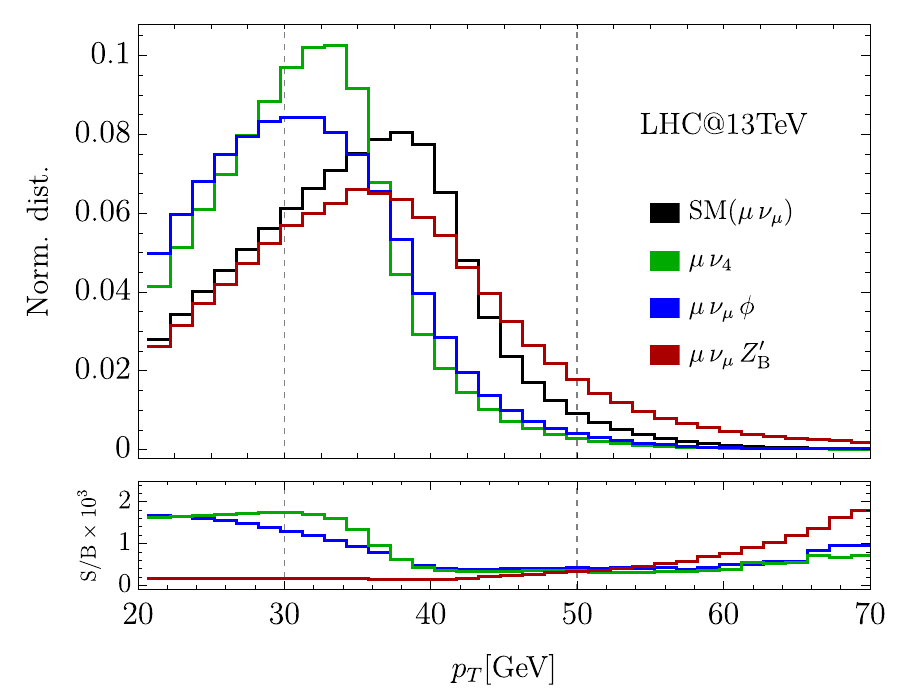}
\end{minipage}
\begin{minipage}{0.48\textwidth}
\centering
\includegraphics[width=1\textwidth]{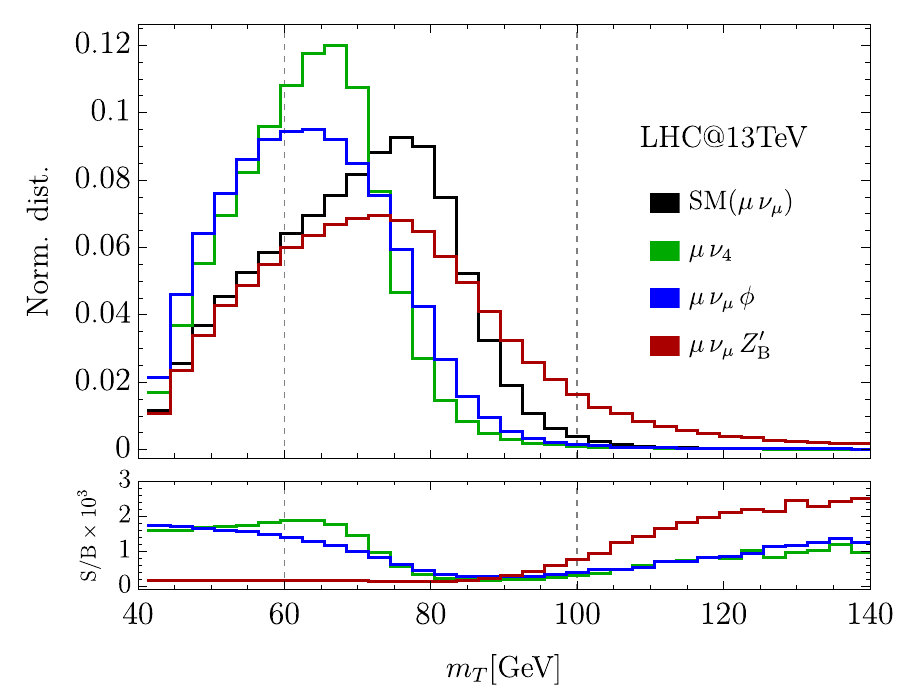}
\end{minipage}
\caption{\footnotesize Normalized kinematic distributions for the models presented in Sec.~\ref{Sec:Neut},~\ref{Sec:Massv}, and Sec.~\ref{Sec:CatB}. The reference points in the new physics parameter space are $(\lambda_{\mu \mu} =1 ,\, m_{\phi} = 10\text{ GeV})$ for the ``neutrinophilic scalar'' (blue lines), $(|U_{\mu 4}| =0.04 ,\, m_{\nu_4} = 30\text{ GeV})$ for the ``heavy neutrino'' (green lines), and $(g_{Z'} ,\, m_{\zprime} = 10\text{ GeV})$ for the hadrophilic $\zprime$ (red lines). Vertical dashed lines delimit the region used by ATLAS for its latest $W$ boson mass measurement~\cite{ATLAS:2017rzl,ATLAS-CONF-2023-004}. The kinematic ranges shown here correspond to the sub-electroweak region shown in Fig.~\ref{fig:Cartoon}.}
\label{Fig:Dist}
\end{figure}

\section{Anomalous $W$ decay}\label{Sec:CatA}
In this section, we complete the analysis of \cite{Agashe:2023itp} and present two additional scenarios that modify the $W$-boson decay. 

\subsection{Emission of a neutrinophilic scalar}\label{Sec:Neut}

The first example that we study is the ``neutrinophilic'' scalar model of \cite{Berryman:2018ogk,de_Gouv_a_2020,Dev:2021axj}. The new ingredient of this model is an effective interaction between the SM and a new ``invisible'' scalar $\phi$ of the form 
\begin{align}\label{Eq:NeutrinoHiggs}
 \mathcal{L} \supset \frac{c_{\alpha \beta}}{\Lambda^2} \left( L_{\alpha} H \right) \left( L_{\beta} H \right)\phi \,. 
\end{align}
This new particle $\phi$ is a SM singlet with mass $M_{\phi}$ and it is assumed to be lighter than the Higgs boson. At low energies, after electroweak symmetry breaking, Eq.~\eqref{Eq:NeutrinoHiggs} produces a new interaction between the SM neutrinos and this new scalar
\begin{align}\label{Eq:NeutrinophilicNeutrino}
 \mathcal{L}_{\rm int} \supset \sum_{\alpha, \beta = e, \mu, \tau} \frac{1}{2}\lambda_{\alpha \beta} \nu_\alpha \nu_\beta \phi\,,
\end{align}
where $\lambda_{\alpha\beta}$ parameterizes the effective coupling of $\phi$ to neutrinos. 
In fact, Eq.~\eqref{Eq:NeutrinophilicNeutrino} opens the new decay channel via radiative emission of $\phi$ from $\nu$ 
\begin{align}
 W \to \ell\, \nu^{*} \rightarrow \ell \, \nu \, \phi \,.
\end{align}
Since $\phi$ is invisible by construction, this decay mode gives $\ell+\met$ (see Fig.~\ref{Fig:NeutrinophilicScalar}), directly affecting the $W$ boson mass measurement. 
We refer to \cite{Berryman:2018ogk,de_Gouv_a_2020,Dev:2021axj} for any general aspects of the phenomenology and the modeling of this scenario. Here, we simply focus on the sensitivity of the $W$ boson mass measurement to the interaction in Eq.~\eqref{Eq:NeutrinophilicNeutrino}.

We show in Fig.~\ref{Fig:Dist} the $p_T^\ell$ and $m_T$ distributions of this anomalous decay (blue lines), compared to the SM (black lines) for a benchmark point in the new physics parameter space. As expected, the net effect is that both distributions become softer, populating the region below the Jacobian peak from on-shell $W$ production. The softening of these spectra is expected, as it is unavoidable that a certain fraction of the $W$ energy is taken away by $\phi$. For this analysis, we thus focus on the sub-electroweak strategy, mostly following the ATLAS \cite{ATLAS:2017rzl,ATLAS-CONF-2023-004} and CDF \cite{CDF:2022hxs} analyses. We depart from the exact selection of ATLAS and CDF in that we slightly enlarge the analysis ranges of each observable (see Tab.~\ref{tab:CUTS} for details). The use of slightly larger analysis ranges is not crucial to bound new physics with our method, but there is a clear improvement from using the proposed analysis ranges in our analysis described in Sec.~\ref{subsec:technicalities}. We remark that in this analysis, we use the same recoil $u_{T}$, thus leaving the QCD aspects of the analysis substantially unchanged. Most of the gain comes from including softer $m_{T}$ and $p_{T}^{\ell}$, which appears safe from the systematics point of view. In any case, the final assessment of the optimal range for this search needs to be determined by the experiments themselves.

\begin{figure}[t]
\centering
\begin{minipage}{0.48\textwidth}
\centering
\includegraphics[width=1\textwidth]{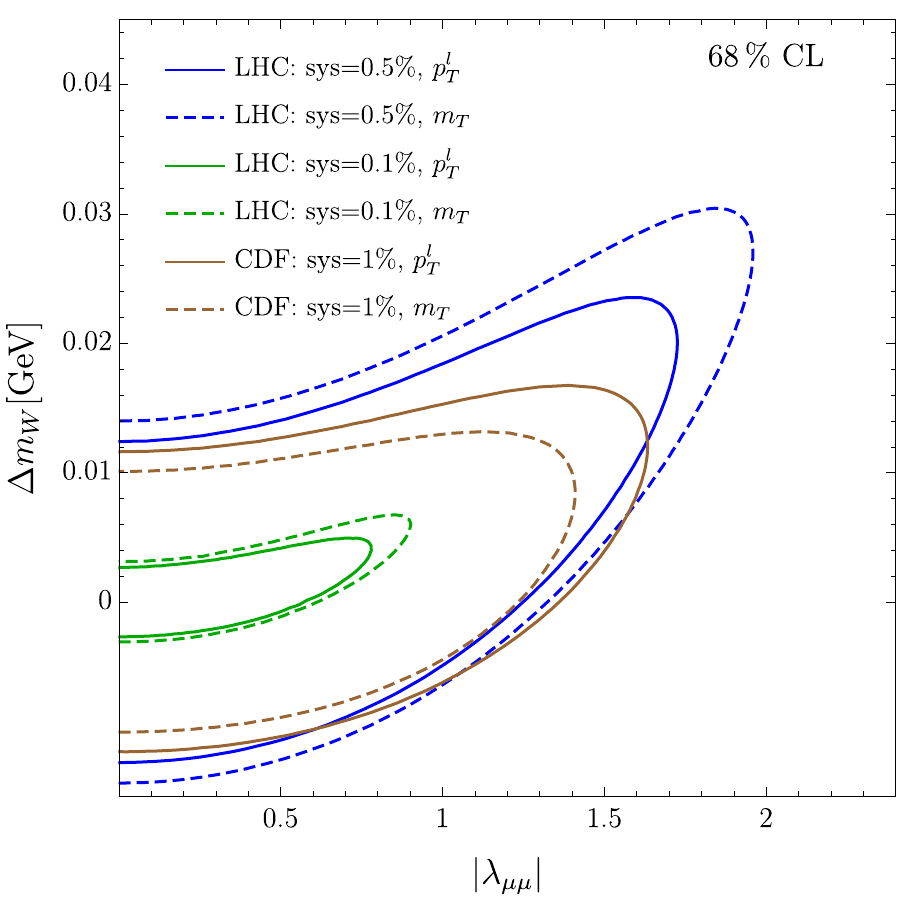}
\end{minipage}
\begin{minipage}{0.48\textwidth}
\centering
\includegraphics[width=1\textwidth]{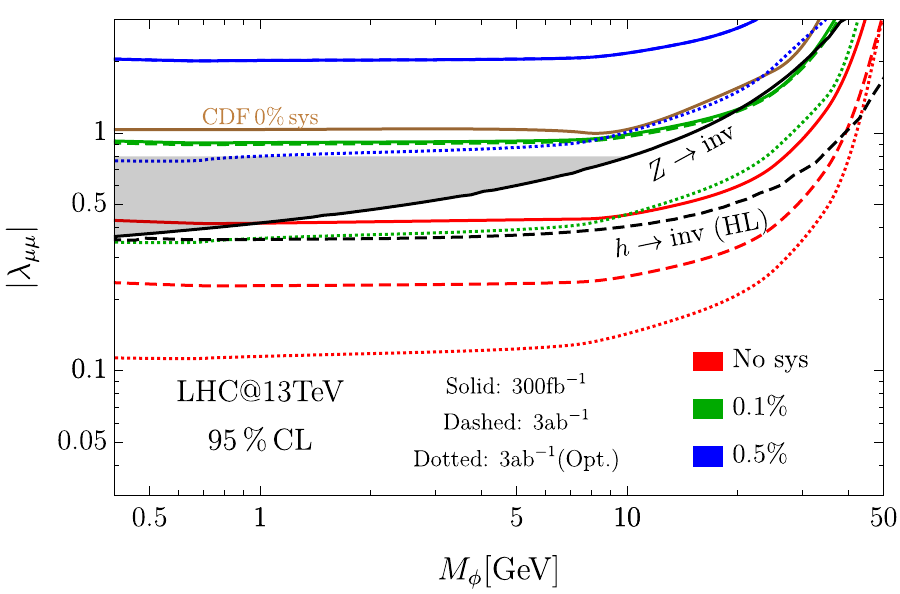}
\end{minipage}
\caption{\footnotesize Sensitivity projections for the neutrinophilic scalar model. Left panel: 68$\%$ CL reach on the $(|\lambda_{\mu \mu}|, \Delta m_W)$ plane with $M_{\phi} = 10 $ GeV. 
Right panel: 95$\%$ CL reach on the $(M_{\phi}, |\lambda_{\mu \mu}|)$ plane. Current constraints from invisible $Z$ decays and projected constraints from invisible Higgs decays are taken from \cite{de_Gouv_a_2020}. In the constraint from invisible $Z$ decays, the gray band gives a rough measure of the uncertainty related to higher order corrections. For the solid and dashed lines of our projections we adopt the benchmark setup described in Sec.~\ref{subsec:technicalities} and we marginalize on $m_W$. The dotted lines assume an ``optimistic'' setup and set $\Delta m_W = 0$. See main text for details. \label{Fig:Neutrino}}
\end{figure}

Before describing the sensitivity projections from the sub-electroweak strategy to this model, we first investigate the impact of this scenario on the determination of $m_W$.
We focus on the $\mu+\met$ final state. In principle, the electron final state can be equally sensitive, but we do not pursue it as it involves different detector effects.
We fix the value of the new scalar mass to $M_{\phi} = 10$ GeV and study the $\chi^2_{\mathcal{O}}$ in Eq.~\eqref{Eq:ChiSqu1} in the $(|\lambda_{\mu \mu}|, \Delta m_W)$ plane. 
The result is reported in the left panel of Fig.~\ref{Fig:Neutrino} for $m_{T}$ and $p_{T}^{\ell}$, LHC and Tevatron colliders, and several possible choices of systematics. 
For this analysis, we adopt the setup and the parameters described in Sec.~\ref{subsec:technicalities} and Tab.~\ref{tab:CUTS}. 
Here we do not report any additional constraints on the new physics parameter space, aiming to show the correlation between $m_W$ and the latter. 

Remarkably, there is a nontrivial interplay between the best-fit value for $m_W$ (via $\Delta m_W$) and $|\lambda_{\mu \mu}|$. For a non-zero $|\lambda_{\mu \mu}|$, $\Delta m_W$ slightly prefers a positive value. That can be understood by noticing that three-body decay makes the $p_T^\ell$ spectrum softer compared to the SM. 

Assuming ``SM-like'' data with a given $\overline{m}_W $, the best fit prefers a shift toward $m_W>\overline{m}_W$ ($\Delta m_W>0$), as a compromise to partially compensate for the softening of the $p_T^\ell$ spectrum from the three-body decay. 
This is completely analogous to what we have found for the invisibly decaying $L_{\mu}-L_{\tau}$ scenario in \cite{Agashe:2023itp}. We remark that this effect is present for both $p_T^\ell$ and $m_T$, thus seems not to be linked to pile-up and detector responses, which are very different for these two distributions. 
Comparing results for different degrees of systematics, the effect does not change significantly, as can be observed in Fig.~\ref{Fig:Neutrino}. A similar effect is observable, but quantitatively milder, for the CDF. A main difference between the CDF and ATLAS selections is related to the fact that CDF requires smaller hadronic activity (see Tab.~\ref{tab:CUTS}). This results in sharper kinematic distributions\footnote{In the limit that the hadronic recoil $\vec{u}_T \rightarrow 0$, $W$ bosons are produced at rest, and the resulting kinematic distributions $p_T^\ell$ and $m_T$ have a sharp Jacobian peak with a steep edge.} on which it is harder to hide the soft lepton contribution for the three-body decay of this model.

In the right panel of Fig.~\ref{Fig:Neutrino}, we give the sensitivity projections on the $(M_\phi, |\lambda_{\mu \mu}|)$ plane. The 95\% CL value for $ |\lambda_{\mu \mu}|$ is obtained for each $M_{\phi}$ from a combined $\chi^{2}$ summing independent $\chi^{2}$ values of the form Eq.~\eqref{Eq:ChiSqu1} for each distribution. For LHC, we minimize $\chi^{2} = \chi^{2}_{{m_{T}}} + \chi^{2}_{{p_{T}^{\ell}}}$, and for CDF we minimize $\chi^{2} = \chi^{2}_{{m_{T}}} + \chi^{2}_{{p_{T}^{\ell}}} + \chi^{2}_{{p_{T}^{\text{miss}}}}$. In the same plane, we also report the main constraints from other searches. In particular, masses below $M_{\phi} \lsim 0.5 $ GeV are heavily constrained by meson decay spectra \cite{de_Gouv_a_2020}. We omit this bound by only showing the parameter space for $M_\phi >$ 0.5 GeV. Larger masses are well tested by the precise measurement of the $Z$-boson invisible decay width and will be better tested by invisible Higgs decays. The latter will be mostly effective at the High Luminosity stage of the LHC. Additional constraints on the region of interest are expected from DUNE and by a direct test of the interaction in Eq.~\eqref{Eq:NeutrinoHiggs} (see \cite{de_Gouv_a_2020} for a comprehensive list of constraints). These cover only a few corners of the parameter space in Fig.~\ref{Fig:Neutrino}, hence we omit them for the sake of clarity of the figure. Regarding the bound from $Z$-boson decays to invisible extracted from \cite{de_Gouv_a_2020}, in Fig.~\ref{Fig:Neutrino} we report a band rather than a line. The reason is that in \cite{de_Gouv_a_2020} no loop corrections to $Z \rightarrow \nu \nu$ decays, induced by loops of $\phi$, are considered. Consequently, the $Z$ decay to invisible gets an unphysical divergence for low masses $M_{\phi}$ of the form $\log \frac{m_Z}{M_{\phi}}$~\cite{Berryman:2018ogk} and this strengthens the corresponding bound. Thus, we obtain the band in Fig.~\ref{Fig:Neutrino} from the line of \cite{de_Gouv_a_2020} and assuming the corresponding bound to become insensitive to $M_{\phi}$ below 10 GeV. We believe that this provides a conservative, yet realistic, estimate.

In Fig.~\ref{Fig:Neutrino}, we report 9 different lines corresponding to the expected bounds from the proposed analysis for variations of systematics, statistical treatment, and assumed luminosity. Specifically, for all the solid and dashed lines, we follow the benchmark setup described in Sec.~\ref{subsec:technicalities} and Tab.~\ref{tab:CUTS} with various luminosities and systematics indicated in the figure. Furthermore, aiming at a conservative result, we obtain these lines marginalizing over $\Delta m_W$. We also follow the same procedure for the CDF projections. The dotted lines show ``optimistic'' projections for our methods applied at the HL-LHC. These projections are valid for different degrees of systematic uncertainties,
negligible effect of pile-up, finer bin size of $0.25$ GeV for $p_T^{\ell}$ and $0.5$ GeV for $m_T$, and, finally, assuming prior knowledge of $m_W$, i.e., fixing $\Delta m_W = 0$. 

All in all, Fig.~\ref{Fig:Neutrino} suggests that the same data currently employed for the $W$ boson mass measurement can be repurposed to provide a competitive probe of this model. As in many cases with precision physics analyses, this method can yield very remarkable results, complementing and even superseding the other proposed probes of the neutrinophilic scalar if one keeps the systematic uncertainties under good control.

\subsection{Two-body decay into a heavy neutrino}\label{Sec:Massv}

Another scenario to which our sub-electroweak new physics search strategy is sensitive is the so-called ``heavy neutrino'' scenario \cite{deGouvea:2015euy}. The main novelty is that the {\it three} active SM neutrinos in the flavor basis are a linear combination of {\it four} mass eigenstates, that we denote $\nu_{1,2,3}$ for the massless neutrinos and an additional sterile ``heavy'' neutrino $\nu_{4}$ of mass $m_{\nu_4}$. The active neutrinos are defined as superpositions
\begin{align}\label{Eq:MixingNeut}
 \nu_{\alpha} = \sum_{i=1}^{4} U_{i \alpha} \nu_i\,,\quad \text{ for } \quad \alpha=e,\, \mu,\, \tau\,.
\end{align}
This framework is typical, for instance, of models where the heavy neutrino acts as the portal to the dark sector \cite{Bertoni:2014mva,Batell_2018}. 
We do not elaborate on this aspect any further and instead focus on the collider phenomenology of models of this sort. 
The following interactions induced by the mixing in Eq.~\eqref{Eq:MixingNeut} are of interest
\begin{equation}
 \mathcal{L}_{\rm int} \supset \sum_{\ell = e,\mu,\tau} U_{4\ell} W^+_\mu \bar{\nu}_{4} \gamma^\mu \ell^-_L + \text{h.c.}\,.
\end{equation}
Assuming that the heavy neutrino is effectively invisible, one can see that the above interactions allow an anomalous $W$ decay (see Fig.~\ref{Fig:HeavyNeutrino})
\begin{align}
 W \rightarrow \ell\, \nu_4\,, \label{eq:decay2nu4}
\end{align}
which contributes to the $\ell + \met$ final state.

The resulting $m_{T}$ and $p_{T}^{\ell}$ distributions are shown in Fig.~\ref{Fig:Dist} (green lines). 
Both spectra get softer than what is expected in the SM, as it is expected on general ground for models that affect the decay of real $W$ bosons. 
We therefore slightly extend the analysis ranges compared to the ATLAS and CDF measurements, as to include softer $p_{T}^\ell$ and $m_{T}$. In Tab.~\ref{tab:CUTS} we give the exact ranges that we find to maximize the sensitivity to this model from the analysis described in Sec.~\ref{subsec:technicalities}. As in the case of the neutrinophilic scalar, the use of slightly larger analysis ranges is not crucial to bound new physics with our method, but again a clear improvement is expected by using the proposed analysis ranges. As we have commented already, systematic uncertainties play a great role in our strategy thus a final assessment of the optimal range for this search needs to be determined by the experimental collaborations.
 
 \begin{figure}[t]
\centering
\begin{minipage}{0.48\textwidth}
\centering
\includegraphics[width=1\textwidth]{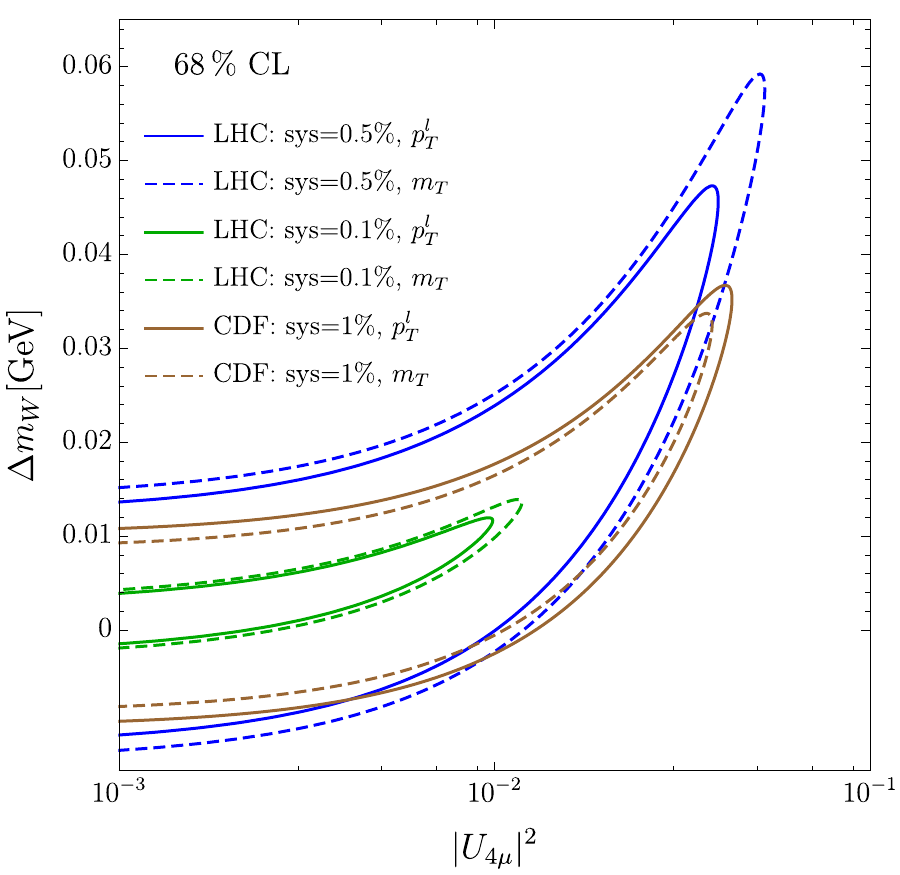}
\end{minipage}
\begin{minipage}{0.48\textwidth}
\centering
\includegraphics[width=1\textwidth]{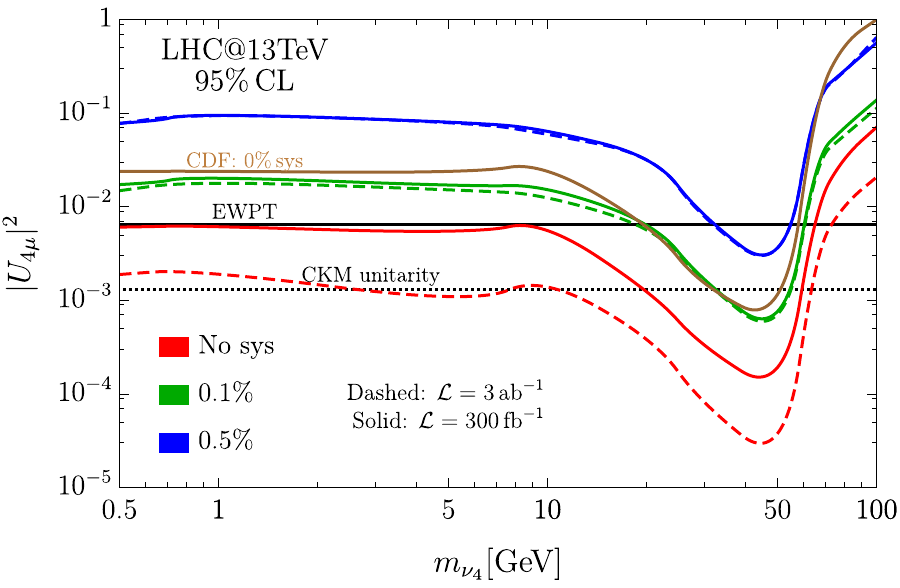}
\end{minipage}
\caption{\footnotesize Sensitivity projections for the heavy neutrino scenario. Left panel: 68\% CL reach on the ($|U_{4\mu }|^2,\Delta m_W $) plane for $m_{\nu_4}=10$~GeV. Right panel: 95\% CL reach on the ($m_{\nu_4} , |U_{4\mu }|^2$) plane. Present constraints are extracted from \cite{Batell_2018}. Our constraints are expected to be roughly the same for electrons ($U_{4e}$), where the CKM unitarity bound does not apply.}
\label{Fig:Massv}
\end{figure}
 
In Fig.~\ref{Fig:Massv} we present our results for this scenario. 
We focus on the $\mu+\met$ final state. In principle, the electron final state can be equally sensitive, but we do not pursue it as it involves different detector effects.
The impact of this model on the determination of $m_W$ on the $(|U_{4 \mu}|^2, \Delta m_W)$ plane is presented in the left panel of Fig.~\ref{Fig:Massv}. 
The figure is obtained taking the baseline setup described in Sec.~\ref{subsec:technicalities} and setting $m_{\nu_4} = 10$ GeV and studying each $\chi^{2}_{\mathcal{O}}$ as indicated by the labels in the figure.
Interestingly, a nontrivial correlation between the mixing $U_{4 \mu}$ and $\Delta m_W$ exists in both CDF and LHC projections, which shows again how new physics can effectively bias the extraction of $m_W$. Yet, we see that the overall effect in this scenario is quite mild, considering the values of $U_{4 \mu}$ for which we find a sizeable $\Delta m_W$ shift are well tested by other experiments, for which we refer to the right panel of Fig.~\ref{Fig:Massv}.

Sensitivity projections for our method and other experiments, {taken from \cite{Batell_2018},} are presented on the $(m_{\nu_4},|U_{4 \mu}|^2)$ plane in the right panel of Fig.~\ref{Fig:Massv}. We stress that for this analysis we employ the baseline setup of Sec.~\ref{subsec:technicalities} and give bounds on $U_{4\mu}$ marginalizing on $\Delta m_W$. This gives conservative bounds on this new physics mixing parameter. 
Our bounds complement nicely the existing ones. Specifically, in the sub-GeV region, the most stringent constraints come from high-intensity meson decay studies. We do not show this bound since Fig.~\ref{Fig:Massv} considers $m_{\nu_4} > $ 0.5 GeV. In the region from few GeV to $m_W$, main constraints are either from electroweak precision tests or from the test of the CKM unitarity \cite{Batell_2018} in the same ballpark as the projected constraints from our method. For heavier $\nu_{4}$ our bounds are particularly effective as the change in the shape of the kinematic distributions is most evident. 
We remark that CKM unitarity bounds apply only to the muonic final state, as they come from meson decays, which produce mostly muon flavor. Bounds from our method can be applied equally well to electrons and muons. Therefore, we expect that the constraints from the measured spectrum of the $m_W$ distribution can be particularly competitive for $\nu_{4}$ coupled mostly to electrons.

\section{Anomalous $W$ production: Hadrophilic $Z'$}\label{Sec:CatB}

As an example of anomalous $W$ boson production, we consider the case of the so-called hadrophilic $Z'$, i.e., a new gauge boson coupled to the baryon number. The effective interaction is described by
\begin{align}\label{Eq:DarkZprime}
 \mathcal{L}_{\rm int} = g_{Z_B'} \bar{q} \slashed{Z}_B' q \,,
\end{align}
with $q$ spanning over all the SM quarks.\footnote{We remark that as the baryon number has an anomalous current, Eq.~\eqref{Eq:DarkZprime} is to be understood only as effective theory supported by an additional Wess-Zumino term. As pointed out, for instance, in \cite{Dror:2017ehi}, this may lead to very strong constraints. Yet these effects are not completely model-independent, as discussed, for instance, in \cite{Dror:2017ehi,DiLuzio:2022ziu}, and we, therefore, do not pursue any further discussion.} We denote the mass of this additional gauge boson by $M_{\zprime}$. Interactions of the $\zprime$ of this form have been widely studied in the literature, either in the context of a simple extension to the SM or as a possible portal to dark matter (see for instance \cite{Carone:1995pu,FileviezPerez:2010gw,Dror:2017ehi,Ilten:2018crw,Boos:2022pyq} and references therein). Like the previous models, we only focus on the collider phenomenology of Eq.~\eqref{Eq:DarkZprime}, under the assumption that the $\zprime$ or its decay products do not leave observable traces in collider experiments, hence the $\zprime$ contributes to the $\met$.

This model belongs to category (B), since the SM $W$ boson can be produced in association with this gauge boson as shown in Fig.~\ref{Fig:Hadrophilic}:
\begin{align}
 p p \rightarrow W (\ell \nu)\, \zprime\,, \notag
\end{align}
which produces a $\ell+\met$ signature. As anticipated, the resulting kinematic distributions give harder spectra than those in the SM, which are shown in Fig.~\ref{Fig:Dist} for $p_T^\ell$ and $m_T$. This motivates us to slightly extend the analysis ranges {toward higher $p_T^\ell$ and $m_T$} compared to the existing ATLAS measurements (see Tab.~\ref{tab:CUTS}) in the sub-electroweak region. This model could also be studied in the circa-electroweak region, but for this paper, we only focus on the sub-electroweak strategy aimed at showing the interplay between the $W$-mass measurement and relatively light new physics. In the high-end part of these extended ranges, the SM production gets less and less copious for larger and larger values of $p_T^\ell$ and $m_T$. This makes the sensitivity of CDF drop significantly compared to previous new physics models. Therefore, we do not include CDF in this analysis. 

To determine the sensitivity of the observables used in the ${W}$ mass measurements, we study the $\chi^{2}=\chi^{2}_{m_{T}}+\chi^{2}_{{p_{T}^{\ell}}}$ computed as described in Sec.~\ref{subsec:technicalities}. At variance with the models studied above, we explicitly simulate only the final state $\mu+\met$ and take into account electrons as a doubling of the rates in each bin. This gives a fair estimate of the sensitivity of our method in the case in which systematics are negligible. This is quite accurate as most of the sensitivity in this model comes from the extended analysis ranges described above. 

Unlike in the previous models, the emission of $\zprime$ can affect other processes, including the production of $Z$ bosons, which is used as calibration for the MC used to obtain data-driven predictions, and especially for the reduction of systematic uncertainties in \cite{ATLAS:2017rzl}.
The $\zprime$ can in principle impact the calibration of MC prediction, and can thus remove part or all the sensitivity of the $\ell+\met$ channel. Such an effect can be mitigated by imposing a $\met$ cut in the selection of the events used for the calibrating from the process $pp\to Z \rightarrow \ell \ell $ that can effectively remove $pp \to Z \zprime$. 
In the following, we do not provide a detailed discussion about this effect, as it is beyond the scope of our work. Rather, we only emphasize that such an effect should be assessed appropriately by the experimental collaborations when implementing the analysis for this model. 

\begin{figure}[t]
\centering
\begin{minipage}{0.48\textwidth}
\centering
\includegraphics[width=1\textwidth]{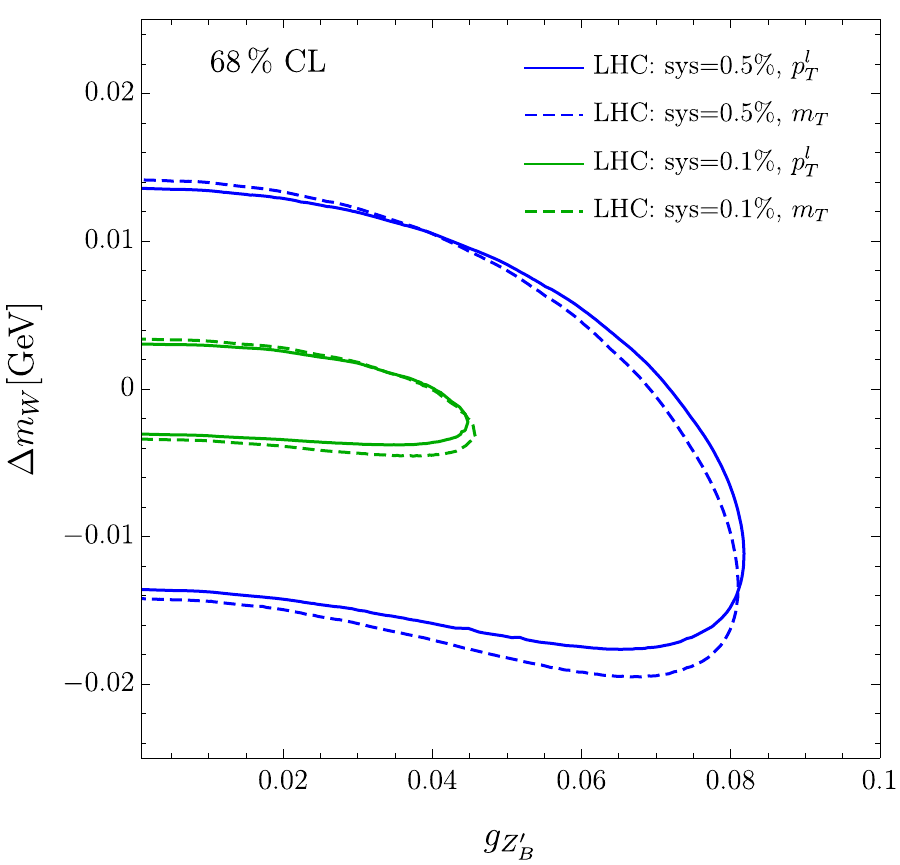}
\end{minipage}
\begin{minipage}{0.48\textwidth}
\centering
\includegraphics[width=1\textwidth]{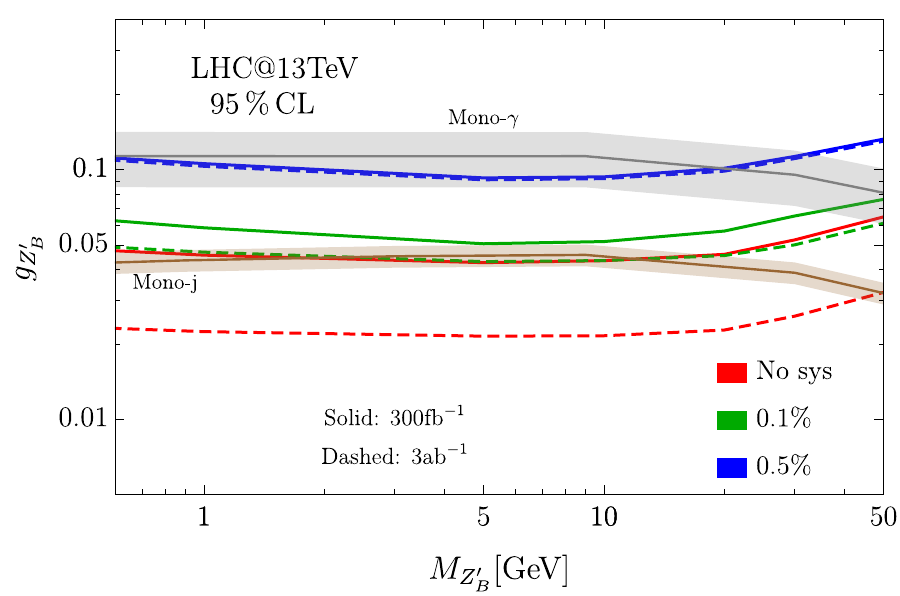}
\end{minipage}
\caption{\footnotesize Sensitivity projections for the hadrophilic $Z'$ scenario. Left panel: 68\% CL reach on the ($ g_{Z_B'}, \Delta m_W$) plane with $M_{Z_B'}$ set to be 10 GeV. Right panel: 95\% CL reach on the ($M_{Z_B'} , g_{Z_B'}$) plane. Existing constraints from mono-jet and mono-photon are obtained by recasts of \cite{CMS:2021far,ATLAS:2020uiq} (see the text for more details).\label{Fig:Hadro} }
\end{figure}

Under the provisions explained above we show the projected sensitivity to the hadrophilic $\zprime$ in Fig.~\ref{Fig:Hadro}. The impact of this model on the $m_{W}$ measurement is reported in the ($g_{\zprime}$, $\Delta m_W$) plane, for a fixed value $M_{\zprime} = 10$ GeV. Interestingly, also in this model there is some correlation between the preferred value for $\Delta m_W$ and $g_{\zprime}$. In contrast to the previously discussed scenarios with modified decays (see Figs.~\ref{Fig:Neutrino} and \ref{Fig:Massv}), the preferred mass difference $\Delta m_{W}$ is negative. The possible effect on the $m_{W}$ value appears to be limited to {\it few} MeV.

The sensitivity projections on the ($M_{Z_B'}, g_{Z_B'}$) plane are presented in the right panel of Fig.~\ref{Fig:Hadro} together with constraints on this scenario from other searches. The main constraints arise from the Run-2 mono-photon and mono-jet searches at the LHC~\cite{CMS:2021far,ATLAS:2021kxv,ATLAS:2020uiq} with an integrated luminosity 138 fb$^{-1}$. These searches report their results in the context of a simplified model in which a new gauge boson $A'$ (dubbed ``mediator'') couples to dark matter $\chi$ and to the quarks. We must adapt the bounds provided in this model to fit our needs. In particular, there are no public results for a new gauge boson lighter than 50~GeV decaying into dark matter.\footnote{We stress that the absence of these light new physics spectra in the publicly available limits is a delicate point. As a matter of fact, we find that these spectra have significantly relaxed bounds on the new physics coupling, thus potentially indicating a loophole of the standard searches.}
Therefore, we recast the results of these analyses to estimate a bound on the coupling $g_{\zprime}$ in the region of interest for our work $ M_{\zprime} \in [0.5, 50]\GeV$, assuming $M_\chi = M_{\zprime}/3$. For lighter $\zprime$ we expect our recast to not be sensitive to the exact mass $M_{\zprime}$. Lighter $M_{\chi}$ would also not significantly change our recast, as the kinematics of the decay $\zprime \to \chi \chi$ is controlled by $\beta^{2}={M^{2}_{\zprime}}/{ ( 4 M^{2}_{\chi}) }$.

For our recast, we simulate at LO the process $pp \to A'+X$ for $X=jet, \gamma$ with \textsc{MadGraph5\_aMC@NLO}~v3.42~\cite{Alwall:2014hca} using a UFO~\cite{Degrande:2012aa} implementation of the model \textsc{DMsimp\_s\_spin1 } used by the experiments~\cite{Albert:2017ab,Boveia:2016ys}.
We pick as recast bound the strongest bound obtained among the several inclusive signal regions considered by the relevant experimental works~\cite{CMS:2021far,ATLAS:2021kxv,ATLAS:2020uiq}. We checked the accuracy of this method for choices of $M_{A'}$ and $M_{\chi}$ as close as possible to our case and we find that we reproduce the results of the mono-photon search up to a factor 25\% in the coupling $g_\zprime$. For the mono-jet recast, we need to adjust our result, strengthening the constraint by a factor of 2. We observe that these adjustment factors are quite close to a constant for mono-jet, within 10\% in our set of spectra, and have a stronger variability than for mono-photon. Thus, we consider our mono-jet recast to have a 10\% uncertainty due to the lack of accuracy of our recast. For the mono-photon bound, we do not need to adjust the results, and we conservatively ascribe a 25\% uncertainty. The origin of these mismatches is not investigated further, as we intend to only obtain a ballpark value for the bound that the experimental collaboration might put on this scenario if they use traditional mono-X searches.

The bound from the recasting of the mono-jet and mono-photon searches and the bounds that we find with our method are shown in the right panel of Fig.~\ref{Fig:Hadro}. All in all, we find that the recast mono-jet bounds are comparable with those from the precision study of the $\ell+\met$ final state. We remark that for our method we present bounds on the new physics coupling marginalizing on $\Delta m_W$ in our $\chi^{2}$, which results in conservative bounds on new physics. 

It is worth noting that, even for small systematic uncertainties around $0.1\%$, improvements are to be expected going from the LHC to the HL-LHC. This behavior is not observed in the previous analyses in Sec.~\ref{Sec:CatA}, where the reach is dominated by systematics already at the LHC. This difference can be ascribed to the fact that new physics in the decay of the $W$ boson affects the soft part of the distribution, where the highest-rate backgrounds populate the phase-space. On the contrary, in the case of production of $W$ associated with new physics, the effects on the kinematic distributions extend to the regions of phases-space that are not so copiously populated by SM processes, on which the higher statistics of the HL-LHC is expected to bring the most benefits.

\section{$\ell+\met$ not associated with the $W$ boson \label{sec:SUSY} }

\begin{figure}[t]
 \centering
 \begin{minipage}{0.48\textwidth}
 \centering
 \includegraphics[width=1\textwidth]{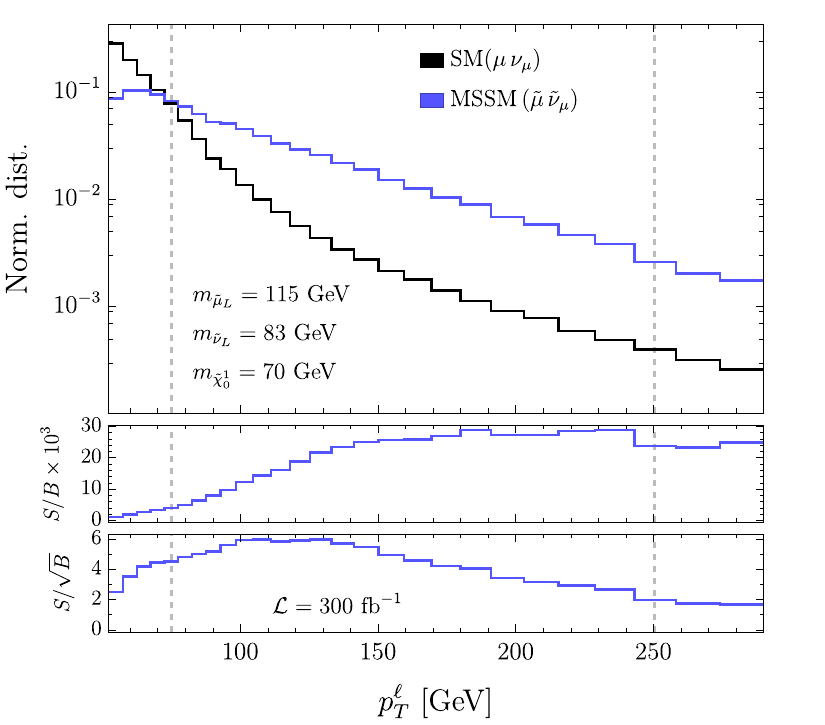}
 \end{minipage}
 \begin{minipage}{0.48\textwidth}
 \centering
 \includegraphics[width=1\textwidth]{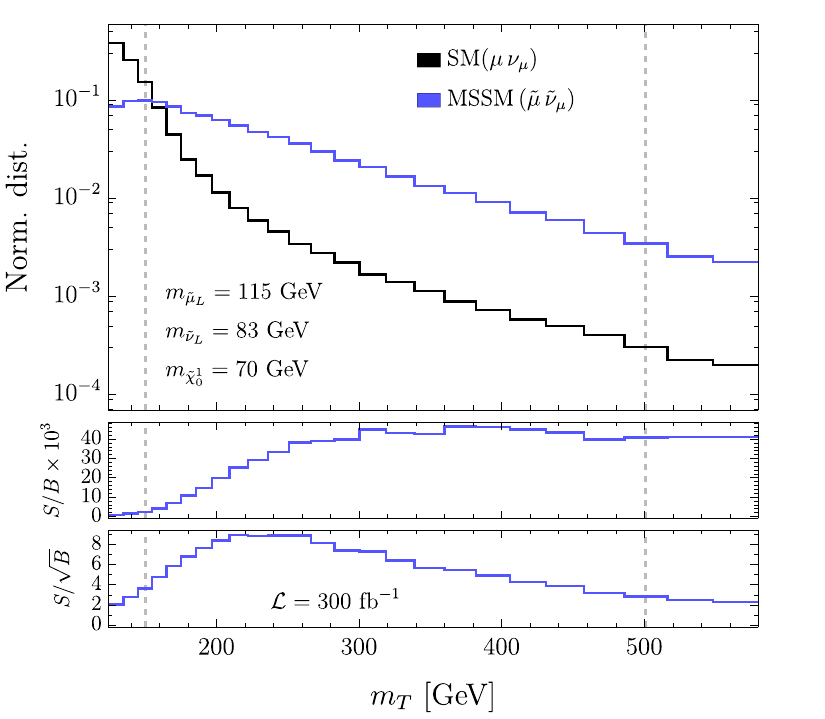}
 \end{minipage}
 \caption{\footnotesize Distributions of SUSY slepton-sneutrino production as signal and Drell-Yan as background for $p_T^\ell$ on the left and $m_T$ on the right. The bottom panels show S/B and S/$\sqrt{B}$ for $\mathcal{L}=300~{\text{fb}}^{-1}$. The selections for both panels are inspired by the CMS $W'$ search and are summarized in Tab.~\ref{tab:susy_cuts}. The binning is described in the text. The vertical dashed lines delimit the range we use for our analysis. The SUSY distributions are taken from a benchmark point of ($m_{\tilde{\mu}_L}$ = 115 GeV, $m_{\tilde{\nu}_L}$ = 83 GeV, $m_{\tilde{\chi}_0^1}$ = 70 GeV). The kinematic ranges shown here correspond to the circa-electroweak region shown in Fig.~\ref{fig:Cartoon}. }
 \label{fig:SUSY_dists_WPrime}
\end{figure}

New physics producing $\ell+\met$ unrelated to on-shell $W$-boson production can be constrained by our method. There are a variety of models belonging to this category, as emphasized in Sec.~\ref{sec:GenSet}. The example that we study is:
\begin{equation}
pp\rightarrow \tilde{\ell}\, \tilde{\nu}_\ell \,, 
\end{equation}
that is slepton-sneutrino production in the MSSM with $\tilde{\ell} \to \ell\, \tilde{\chi}_1^0$ (see Fig.~\ref{Fig:MSSM}). This analysis completes and extends our previous results in \cite{Agashe:2023itp}.

More precisely, in Ref.~\cite{Agashe:2023itp} we focused on the sub-electroweak strategy, showing that the same data responsible for the $W$ boson mass measurement are particularly sensitive to the MSSM compressed region $m_{\tilde{\mu}}- m_{\tilde{\chi}}\lesssim m_W$ via $\tilde{\ell} \,\tilde{\nu}_\ell$. In the present work, we reconsider this channel to assess the sensitivity of data at higher-$p_T$ considering the $circa$-electroweak strategy as well.
The main reason for bringing our attention to this region is that more canonical analyses, e.g., standard $W'$ searches~\cite{CMS:2022krd}, have focused mostly on very large $p_{T}$ values, ending up pursuing new physics in the $supra$-electroweak regime. 

As it can be inferred from the $S/B$ ratio shown in Fig.~\ref{fig:SUSY_dists_WPrime}, valuable sensitivity to new physics arises at moderate $p_{T}$ not far from the typical values used in the analysis devoted to high-precision $m_{W}$ measurements. To pursue such new physics we adopt the selection reported in Tab.~\ref{tab:susy_cuts}. For comparison, we also report in the table the selection used by $W'$ searches of CMS.
We stress that the selection criteria are largely borrowed from the $W'$ search and not optimized to isolate slepton-sneutrino production. However, the survival rate of the SUSY signals is large enough to give us sensitivity to explore the compressed region of the parameter space that is currently beyond the reach of conventional dislepton searches \cite{ATLAS:2022hbt,CMS:2023qhl,ATLAS:2019lff}.

The main difference of our strategy from the $W'$ search is that we pursue a more refined analysis of the shape of the kinematic distributions with finer bins.
For reference, we recall that near the Jacobian peak of the $p_{T}^{\ell}$ distribution, a bin width of 1~GeV is used by CDF and ATLAS. Beyond the Jacobian peak the SM rate decreases, thus larger bins are desirable to keep the statistical uncertainty low enough.
For our analysis, we find a good balance 
using 5 GeV bin width for $p_T^\ell \in [75, 90]$~GeV and exponentially growing bins\footnote{The exact formula to obtain the edges of the $p_T^\ell$ bins is the following: bin edges are given by $b = e^x$, where $x$ ranges from $\ln(p_T^{\text{min}}/\text{GeV})$ to $\ln(p_T^{\text{max}}/\text{GeV})$, with $\Delta x = [\ln(p_T^{\text{max}}/\text{GeV})-\ln(p_T^{\text{min}}/\text{GeV})]/n$, where $n=20$ is the number of bins. This gives $n+1$ bin edges, corresponding to $n$ bins.} for $90 < p_T^\ell < 250$ GeV. 
For $m_T$, the bin widths are twice those of $p_{T}^{\ell}$, as the associated analysis range is doubled from that of $p_T^\ell$. The resulting distributions of $p_T^\ell$ and $m_T$ for MSSM sleptons and SM background $pp \rightarrow \mu \nu$ are shown in Fig.~\ref{fig:SUSY_dists_WPrime}.

\begin{table}[t]
 \centering
 \begin{tabular}{c||c|c||}
 &$W'$ ($\mu$) \cite{CMS:2022krd} & MSSM \\ \hline
 
  $p_T^{\ell}$ minimum (GeV) & $> 50$ & $>50$\\ 
   
  $p_T^{\ell}/p_T^{\text{miss}}$&$>0.4$ \& $<1.5$& $>0.4$ \& $<1.5$\\
   
  $m_T$ minimum (GeV) & $> 120$ & $>120$ \\ 
  
  $\Delta \phi(\vec{p}_T^{\,\; \ell}, \vec{p}_T^{\,\; \text{miss}})$ & $> 2.5$ & $> 2.5$ \\
  
  $p_T^{\ell}$ range (GeV) & $ [50,4000]$ & $[75,250]$ \\
  
  $m_T$ range (GeV) & $ [120,7000]$ & $[150,500]$ \\

  $p_T^{\ell}$ bin width (GeV) & $80$ & $5-14$ \\
  $m_T$ bin width (GeV) & $98$ & $10-28$
  
 \end{tabular}
 \caption{Kinematic cuts, analysis ranges, and bin widths considered for our MSSM slepton-sneutrino analysis, compared to those used in the $W'$ search by CMS.}
 \label{tab:susy_cuts}
\end{table}

The data at large $p_{T}$ is not used in our analysis, as the mass scale of the new physics that we are after implies no new physics effects to be expected there. Indeed, we cap the analysis ranges for our analysis by looking at the $S/\sqrt{B}$ and $S/B$ distributions for the signal of interest, aiming to include the peak of the $S/\sqrt{B}$ and some padding around it, see  for example, Fig.~\ref{fig:SUSY_dists_WPrime}. We remark that our results depend only mildly on the exact location of the end-point of the analysis range around the values indicated in Tab.~\ref{tab:susy_cuts}.

In our projections, we only consider Drell-Yan (DY) $W$ boson production as background, and we do not include other possible sources of SM backgrounds such as single top production, top quark pair production $t\bar{t}$, diboson production $WW$ and $WZ$, $\tau$, and mistagged $Z$. After imposing our cuts, we find that these subdominant backgrounds only contribute less than 4\% of DY and therefore are ignored for the rest of the analysis.

\begin{figure}[t]
 \centering
 \includegraphics[width=0.5\textwidth]{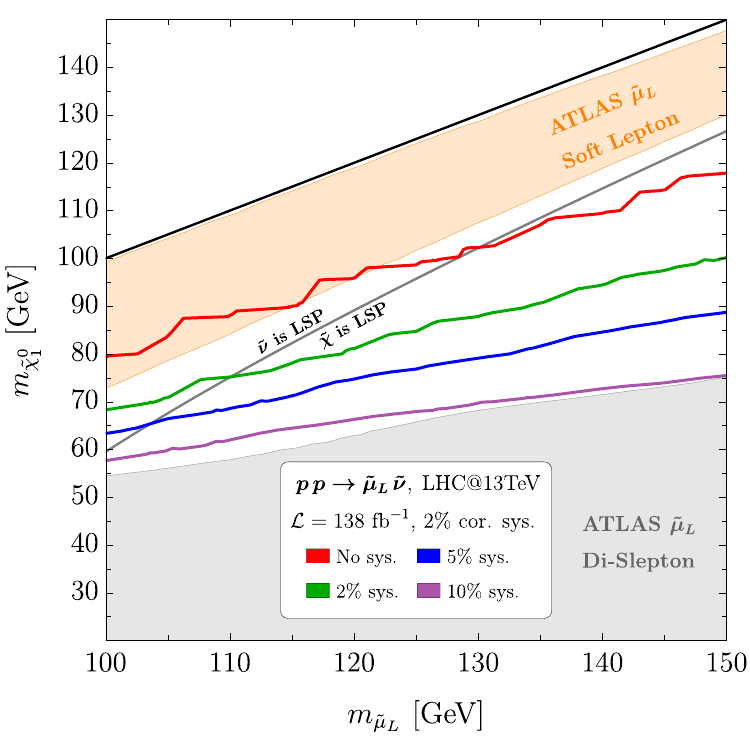}
 \caption{\footnotesize The $95\%$\ CL projected sensitivity to SUSY slepton-sneutrino production at LHC, using $p_T^\ell$ and $m_T$ distributions beyond the Jacobian peak of the SM $W$ boson DY production. The analysis is summarized in Table~\ref{tab:susy_cuts}.}
 \label{fig:SUSY_tail_moneyPlot}
\end{figure}

In the spirit of complementing our work in \cite{Agashe:2023itp}, which was focused on the repurposing the $m_{W}$ measurement, in this work we consider an analysis range that does not include the Jacobian peak of SM $W$ boson production. For this reason, we expect that the sensitivity of this analysis to variations of the measured mass of the $W$ boson is negligible. Therefore, when obtaining the $\chi^2$ for a choice of SUSY parameters, $m_W$ is not a marginalized parameter. Instead, $m_W$ is fixed to the measured ATLAS value throughout. The $\chi^2$ that we study is $\chi^2_{p_{T}}+\chi^{2}_{m_{T}}$. We compute its values comparing the LO theory SM prediction, assumed to be the measured data, with templates in which we vary the slepton and neutralino masses $m_{\tilde{\ell}}$ and $m_{{\tilde{\chi}_{1}^{0}}}$. The sneutrino mass is a dependent parameter, that we obtain from the slepton mass saturating the $D$-term relation in the MSSM \cite{Martin:1997ns,Agashe:2023itp}. 

In Fig.~\ref{fig:SUSY_tail_moneyPlot}, we show 95\% CL sensitivity projections for the LHC current luminosity 139~$\text{fb}^{-1}$, assuming a 2\% correlated systematics on the luminosity, and additional uncorrelated per-bin systematics as per the color code ranging from 0 to 10\%. 
It is worth noting that the benchmark uncorrelated systematics that we employed are quite smaller compared to the $W'$ search~\cite{CMS:2022krd}. Our choice seems tenable, however, because we focus our analysis on the region between $200-400$ GeV, whereas \cite{CMS:2022krd} insists on much larger values of $p_{T}$ and $m_{T}$ on the TeV tail of the kinematic distributions.
In the figure, we also show the part of the plane currently probe by ATLAS searches for MSSM production of smuons. Notably, the method that we propose is sensitive to regions of the slepton-neutralino mass plane not covered by present searches. 

Finally, we remark that the present analysis is independent of the analysis presented in~\cite{Agashe:2023itp}, as the ranges of $p_{T}$ and $m_{T}$ used in the two analyses are disjoint. Thus the present result provides a further complementary new probe to close the gap in the SUSY compressed region, described by our earlier paper.

\section{Conclusions}\label{Sec:Conclusion}
The aim of this paper is to assess the sensitivity of high-precision collider data to new physics. We specifically examine the case of $\ell+\met$ and focus on new physics signals producing typical lepton $p_T$ softer than the EW scale, $\mathcal{O}(100)\GeV$ for which reliable calibration data is available, or slightly harder scale where the SM predictions and detector modeling are still under control. We refer to these two regions as sub-electroweak and circa-electroweak, as illustrated in Fig.~\ref{fig:Cartoon}. Unlike the high-$p_T$ region (referred to as supra-electroweak) which has been heavily exploited in conventional new physics searches, the sub- and circa-electroweak regions are not the typical target for BSM. In this paper we demonstrate that, provided sufficient control of the various sources of uncertainties, at a per-mil level in the sub-electroweak or at a percent level in the circa-electroweak, these regions can serve as optimal probes for relatively light new physics, offering complementary tests to other experiments.

Furthermore, as previously noted in \cite{Agashe:2023itp}, BSM searches conducted in regions heavily overlapping with SM measurements exhibit a notable interplay between BSM and SM physics. Specifically, since the same dataset is used for both BSM searches and SM measurements, it necessitates a simultaneous fit of BSM and SM parameters. Consequently, we observe that new physics can bias the extraction of the SM parameters, as we show for the case of the $m_W$ mass measurement, even if only to a limited amount (see Figs.~\ref{Fig:Neutrino}, \ref{Fig:Massv}, and \ref{Fig:Hadro}).

We find that it is in the {\em sub}-electroweak search that the above-mentioned simultaneous fluctuation of 
SM and NP parameters in the measurements
is a key step for unlocking the discovery potential strategy, thus obtaining a measurement combined with limit-setting. The determination of SM parameters is more decoupled from new physics searches in the circa-electroweak strategy. Still, the circa-electroweak strategy benefits hugely from knowing the SM accurately. In fact, the region of phase-space where new physics is expected to appear, though not overlapping with that used for measurements of SM parameters, is influenced by the tails of SM processes. Thus in both cases, there is a nontrivial meshing of SM measurements and searches for new physics.

The list of possible new physics models to which the precision $\ell+\met$ can be sensitive is vast. Thus we proceed to identify a few models that can serve as axes of the map of the space of models. The models are chosen for their underlying value as models of new physics as well as for their feature to induce specific modification of the kinematic distributions under study.
We present sensitivity projections for (HL-)LHC and CDF to BSM physics belonging to three categories: anomalous $W$-production (Sec.~\ref{Sec:CatA}), anomalous $W$-decay (Sec.~\ref{Sec:CatB}) and new physics producing $\ell+\met$ without the propagation of a (nearly) on-shell $W$ (Sec.~\ref{sec:SUSY}). Each category is illustrated by concrete examples in the corresponding section.
The results in this paper, joint with our previous illustrative results \cite{Agashe:2023itp}, cover a large enough set of possibilities that demonstrates the potential to carry out new physics searches in the sub- and circa-electroweak regions by exploiting the highly ``curated'' data used in precision measurements.

For each scenario, we obtain the following concrete results. We investigate anomalous $W$ boson decays through two examples: the radiative emission of a neutrinophilic scalar (Sec.~\ref{Sec:Neut}) and the two-body decay of the $W$ into $\ell$ plus a heavy neutrino (Sec.~\ref{Sec:Massv}). The generic characteristic of these decays is to make the lepton $p_T$ spectrum softer, populating the sub-electroweak region (see Fig.~\ref{Fig:Dist}), the same region responsible for the $m_W$ measurement. In both these examples, we find a nontrivial correlation between the extracted $m_W$ and the new physics parameters as shown in the left panels of Figs.~\ref{Fig:Neutrino} and \ref{Fig:Massv}. Specifically, we find that a positive $m_W$ shift can partially compensate for the effect of new physics and this might result in a smaller extracted $m_W$ in the presence of the latter. This bias turns out to be smaller at CDF than at LHC.
This is due to a different selection for the hadronic recoil and, to a smaller extent, to a different impact of pile-up. 
For the models considered the possible bias is limited to a fraction of the measurement uncertainty from ATLAS. Still these results highlight that more precise measurements in the future may be hindered by bias due to NP and even cause disagreement between LHC and Tevatron results. Regardless of a possible bias in the measurements, we find that the (HL-)LHC can provide a competitive bound for these models, assuming a setup as close as possible to the present $W$-boson mass measurement (see right panels of Figs.~\ref{Fig:Neutrino} and \ref{Fig:Massv}). 

An example of anomalous $W$ production is presented through the initial state radiation of a relatively light ($1-50$ GeV) hadrophilic $Z'$ (Sec.~\ref{Sec:CatB}). In this case the $W$-boson recoils against the initial state radiation $Z'$, making the $p_T$ spectrum harder. Like the previous example, we find that in the range of interest, the sub-electroweak region can be a sensitive probe for this model. We limit this analysis to the LHC, as CDF statistics is too low to provide a competitive bound. On the one hand, we find that the same dataset employed for the measurement of $m_W$ in the $\ell+\met$ final state can provide a probe competitive to present $Z'$ sensitivity from LHC searches in mono-jet or mono-photon. Yet on the other hand, we find again that this kind of BSM can potentially bias $m_W$ with the opposite trend to what is discussed for the anomalous $W$-decay. Although we have not pursued it in this work, a search in precision data in the circa-electroweak region can be sensitive to this model and could help to disentangle a possible effect on the measurement of $m_W$. 

Finally, we provide a study of new physics populating the circa-electroweak region via slepton-sneutrino production in the MSSM (Sec.~\ref{sec:SUSY}). This case study was already investigated in \cite{Agashe:2023itp}, but there, the analysis was limited to the sub-electroweak region. On the contrary, in the present paper, we show that extending the analysis to the circa-electroweak region provides further sensitivity and improves our earlier result, provided control of the systematic uncertainty at a few percent level can be attained in the circa-electroweak region. We remark that this channel was somehow ignored by present LHC searches and has the potential to cover wholly new regions of parameter space of the MSSM (see Fig.~\ref{fig:SUSY_tail_moneyPlot}). 

With upcoming $\ell+\met$ data from the LHC, we foresee opportunities for new physics searches using new observables, such as a doubly differential distribution in lepton $p_T$ and $\eta$. 
This data could potentially be unfolded to the particle level for wider accessibility beyond experimental collaborations, or it could be published together with full information on correlated uncertainties on the kinematic distributions, allowing a more careful re-use and limit-setting from the procedure described above.
Furthermore, constraints will strengthen as statistics increase and precision improves at HL-LHC. 

We anticipate that as these methods are perfected, testing new physics with a nearly model-independent method becomes possible, enabling the pursuit of elusive models, such as the so-called ``inert doublet''~\cite{Deshpande:1977rw,Barbieri:2006dq,Cao:2007rm} or other SUSY scenarios~\cite{Carpenter:2020fnh}, which can produce $\ell+\met$ but are too faint~\cite{Kalinowski:2020rmb} for present sensitivity. 

We believe this strategy to be a novel and promising way to look at present and future collider data, ultimately applicable to a variety of BSM searches and SM measurements.

\section*{Acknowledgements}

The authors would like to thank Alberto Belloni,
Javier Montejo Berlingen,
Cecile Sarah Caillol,
Caterina Doglioni,
Rafael Lopes de Sa,
Valentina Dutta,
Sarah Eno,
Yongbin Feng,
Tao Han,
Philip Harris,
Bodhitha Jayatilaka,
Jakub Kremer,
Gustavo Marques-Tavares,
Mario Masciovecchio,
Patrick Meade,
Federico Meloni,
Pier Francesco Monni, Alessandro Vicini and 
Felix Yu for discussions. 
The work of RF is supported in part by the European Union - Next
Generation EU through the MUR PRIN2022 Grant n.202289JEW4.
The work of K.~A., S.~A., L.~R.~and D.~S.~is supported by NSF Grant No.~PHY-2210361 and by the Maryland Center for Fundamental Physics. 
The work of D.~K. is supported by the DOE Grant No. DE-SC0010813. The work of A.~V.~K. is supported by the DOE Grant No. DE-SC0010007.

\bibliography{main}
\bibliographystyle{JHEP.bst}

\end{document}